\documentclass[12pt]{article}

\usepackage{psfig,epsfig}

\newcommand{\be}{\begin{equation}}
\newcommand{\bea}{\begin{eqnarray}}
\newcommand{\ee}{\end{equation}}
\newcommand{\eea}{\end{eqnarray}}
\def\chic#1{{\scriptscriptstyle #1}}

\catcode`@=11
\newcount\@tempcntc
\def\@citex[#1]#2{\if@filesw\immediate\write\@auxout{\string\citation{#2}}\fi
  \@tempcnta\z@\@tempcntb\m@ne\def\@citea{}\@cite{\@for\@citeb:=#2\do
    {\@ifundefined
       {b@\@citeb}{\@citeo\@tempcntb\m@ne\@citea\def\@citea{,}{\bf
?}\@warning
       {Citation `\@citeb' on page \thepage \space undefined}}%
    {\setbox\z@\hbox{\global\@tempcntc0\csname
b@\@citeb\endcsname\relax}%
     \ifnum\@tempcntc=\z@ \@citeo\@tempcntb\m@ne
       \@citea\def\@citea{,}\hbox{\csname b@\@citeb\endcsname}%
     \else
      \advance\@tempcntb\@ne
      \ifnum\@tempcntb=\@tempcntc
      \else\advance\@tempcntb\m@ne\@citeo
      \@tempcnta\@tempcntc\@tempcntb\@tempcntc\fi\fi}}\@citeo}{#1}}
\def\@citeo{\ifnum\@tempcnta>\@tempcntb\else\@citea\def\@citea{,}%
  \ifnum\@tempcnta=\@tempcntb\the\@tempcnta\else
   {\advance\@tempcnta\@ne\ifnum\@tempcnta=\@tempcntb \else
\def\@citea{--}\fi
    \advance\@tempcnta\m@ne\the\@tempcnta\@citea\the\@tempcntb}\fi\fi}
\catcode`@=12
 
\begin{document}

\begin{flushright}
FTUV-00-08-11\\
\end{flushright}
 
\begin{center}
{\Large {\bf On the charge radius of the neutrino}}
\\[2.4cm]
{\large J. Bernab\'eu}, {\large L. G. Cabral-Rosetti},
{\large J. Papavassiliou},
and 
{\large J. Vidal}\\[0.4cm]
{\em 
Departamento de F\'{\i}sica Te\'orica, Universidad de Valencia \\
E-46100 Burjassot (Valencia), Spain}\\[0.5cm]

\end{center}
 
\vskip0.7cm     \centerline{\bf   ABSTRACT}  \noindent

Using the pinch technique we  construct  at one-loop order a  neutrino
charge  radius, which is  finite,  depends neither on the gauge-fixing
parameter   nor on  the   gauge-fixing    scheme  employed, and     is
process-independent.  This definition stems  solely from an  effective
proper photon-neutrino one-loop vertex,   with no reference to box  or
self-energy contributions.  The    r\^ole  of the  $WW$  box  in  this
construction  is critically examined.  In  particular it is shown that
the exclusion  of the effective  $WW$ box from   the definition of the
neutrino charge  radius is not  a matter of convention  but is in fact
dynamically  realized  when  the  target-fermions  are   right-handedly
polarized.  In this way we obtain a  unique decomposition of effective
self-energies,    vertices,  and   boxes,    which separately  respect
electroweak  gauge invariance.  We elaborate  on the tree-level origin
of   the  mechanism    which  enforces  at  one-loop  level    massive
cancellations  among the   longitudinal  momenta appearing   in  the
Feynman   diagrams, and in particular     those  associated with   the
non-abelian character of the theory.   Various  issues related to  the
known connection between the pinch technique  and the Background Field
Method  are further  clarified.   Explicit closed  expressions for the
neutrino charge radius are reported.

\vskip0.7cm

\newpage

\setcounter{equation}{0}
\section{Introduction}

\noindent

The  neutrino  electromagnetic    form-factor  and the   corresponding
neutrino charge    radius (NCR) are  interesting  physical quantities,
mainly  because  their non-vanishing appreciable  
values  could be   attributed to
physics  beyond the Standard Model (SM).  Within  the SM,
one expects   that  an effective photon-neutrino   interaction will be
generated through one-loop radiative   corrections, giving rise   to a
non-zero, albeit small NCR.  However, the direct calculation of this
quantity has been faced   with serious complications 
\cite{N6,N2,N1,N9} 
which,  in turn, can be traced  back to the  fact  that in non-Abelian
gauge   theories   off-shell  Green's functions      depend in general
explicitly on   the   gauge-fixing parameter  (GFP).   Therefore   the
definition of quantities familiar from scalar theories of QED, such as
effective charges and form-factors, is in general problematic.  In the
case of the NCR, various  attempts to define  its SM value through the
one-loop  $\gamma\nu\nu$  vertex   calculated  in the   renormalizable
($R_{\xi}$) gauges   reveal  that  the   corresponding electromagnetic
form-factor $F_1   (q^2,\xi)$ depends  explicitly  on  the  GFP   in a
prohibiting  way.  In particular, even though  in  the static limit of
zero momentum transfer,  $q^2 \to 0$,  the form-factor $F_1 (q^2,\xi)$
becomes independent of  $\xi$,  its first  derivative with respect  to
$q^2$, which corresponds to the classic definition  of the NCR, namely
$\big <r^2_\nu\,  \big>    = -\, 6\,   \frac{\partial  F_{1}(q^2,\xi)}
{\partial q^2} |_{q^2 = 0}$\, , continues to depend on it.

The  problems associated with the definition  of the NCR within the SM
and beyond have  been examined in detail  in  a series of papers,  and
various proposals have been put forth on how to bypass them 
\cite{N4,N3,N7,N8,N12,N11,N13,N14,N15,N10,N20,N5}. 
 The main idea behind these works  has been to exploit judiciously the
GFP-independence  of the  entire  physical  amplitude, and  eventually
define a piece of it which is kinematically akin to an electromagnetic
form  factor,  and is in  addition  endowed with  a  set  of important
physical properties.   In this context the  r\^ole of the box diagrams
containing   two $W$'s has   presented a long-standing  puzzle. It was
recognized in  \cite{DSM}  that this  box  contains in general crucial
GFP-dependent pieces, which,   when  included to  the  NCR  definition
render the answer GFP-independent.  Thus, whereas in this construction
the  ZZ  boxes   have been naturally   discarded,  since  they  do not
participate non-trivially  in the gauge  cancellations of the NCR, the
$WW$ box is instead included in the definition,  because it is important
for  the gauge cancellations.  The  disadvantage  of including the box
however is  that it introduces  a process-dependence, since it depends
explicitly on the quantum numbers of the target fermions; this fact in
turn would prevent the NCR from being regarded as a target-independent
property, intrinsic to the  neutrino-photon interaction.  Even  though
the process-dependent  contributions may become numerically subleading
when studying the low-energy static limit, where the  NCR has a direct
electromagnetic analogue, from the theoretical point of view one would
like to   be able to   extend its  notion  into  that of  an effective
form-factor,  which  would be  GFP-  and process-independent  for {\it
arbitrary} values of the momentum transfer.

The way out  of this  dilemma  has been presented   in
\cite{JPPT}, in  the context of  the pinch technique   (PT) 
\cite{PTCPT,PT,PT2,DS}: one can construct  a  genuine,
target-independent form-factor endowed with  the crucial properties of
gauge-independence, gauge-invariance  and finiteness, for    arbitrary
values  of the    momentum transfer.  The   essential ingredient   for
accomplishing this was the realization that,  when the target fermions
involved   are  considered   to   be    massless, all    GFP-dependent
contributions stemming from box  and vertex-like Feynman diagrams  are
effectively propagator-like, i.e.  have no dependence on the kinematic
properties or  quantum  numbers of  the  initial and final  states, as
boxes and vertices normally do. The precise mechanism which allows the
identification  and extraction of  these propagator-like contributions
from boxes and  vertices  relies  on  the systematic  exploitation  of
fundamental tree-level Ward identities  (WI).  At tree-level  these WI
enforce     crucial  cancellations    between    $s$-and   $t$-channel
contributions   in  the  $W$-pair production   process 
\cite{PP2,PRW}.
Their one-loop  descendant is  a  set of  vast cancellations involving
contributions originating  from  seemingly  dissimilar  Feynman graphs.
Thus,  all gauge dependence  stemming from vertices and boxes cancels
precisely against  the gauge-dependence stemming from the conventional
one-loop self-energies. As   a result, the remaining   GFP-independent
structures retain their initial kinematic identity; in particular one
can  speak in terms  of GFP-independent effective boxes, vertices, and
self-energies.  Thus,   once    the GFP-dependent pieces    have  been
extracted  from the box,   the remaining GFP-independent  ``pure'' box
should not be considered as a part of the resulting form-factor, which
should be entirely determined from the ``pure'' GFP-independent set of
vertex graphs.  Notice  that,  in the  original proposal,  the gamma Z
self-energy constituted part   of  the form-factor
\cite{JPPT}; whereas  this   is clearly a process-independent 
contribution   it became clear  recently \cite{PRW}
that the $\gamma Z$ self-energy should not be  part of the form-factor
either, since its interpretation as part of the electro-weak effective
charge is physically more appealing.

Ii  should be emphasized  that the aforementioned rearrangement of the
S-matrix  into individually GFP-independent effective boxes, vertices,
and self-energies occurs automatically, at  least up to two-loops
\cite{PT2L},  if
one  adheres to the   well-defined set  of  PT  rules,  which  may  be
summarized   as  follows:  Identify  {\it  all}  longitudinal momenta,
i.e. all momenta which can  trigger the aforementioned elementary Ward
identities,  and   track down  the  resulting   algebraically realized
cancellations.  One, but not the  only,  consequence  of  the above
procedure  is   that  all  dependence    on the  GFP   is  effectively
self-energy-like,  and    finally   vanishes since  the    S-matrix is
guaranteed to be  GFP-independent.   Notice however that  longitudinal
momenta   do not   only  originate   from the tree-level   propagators
appearing in the graphs but also from the appropriate decomposition of
the elementary tri-linear gauge-boson vertices
\cite{TH}.  
This decomposition is characteristic   of the PT and  has far-reaching
consequences; in  particular, the resulting effective one-and two-loop
Green's  functions satisfy  QED-like   WI,    instead of the     usual
complicated Slavnov-Taylor identities. In addition, they {\it coincide
}, at  least up to  two-loops \cite{BFMPT,PT2L}, with  the results one
would  obtain if one were  to calculate  the  Green's functions in the
Background Field Method  (BFM) \cite{BFM1,Abb,AGS,PAG}  at the special
value   of the (quantum)  GFP $\xi^{\chic  W}_{\chic  Q} =  1$.  It is
important to  appreciate that the  aforementioned vertex decomposition
has to be  carried out even if  all gauge dependences have  explicitly
cancelled; conversely, it can be  carried out at  the beginning of the
calculation,  before    any  of   the   GFP cancellations   have  been
implemented,  i.e the order  of triggering the  WI and identifying the
various propagator-like pieces is immaterial, as it should.

Even though the correct  interpretation of the  correspondence between
the PT and the BFM at $\xi^{\chic W}_{\chic Q} = 1$ has been discussed
in the literature
\cite{PP2,PRW,PTBFM95}, 
it has often been misinterpreted as constituting an arbitrary choice
rather than a result singled out in a unique way by the full
exploitation of the underlying Becchi-Ruet-Stora (BRS) symmetries
\cite{BRS}, intrinsic to the S-matrix (or any ostensibly
gauge-invariant quantity such as  a  Wilson  loop, for example).    In
addition,  the proposal to exclude the  pure box from  the  NCR on the
grounds  of  the process-dependence   that   it induces  has not  been
implemented when imposing experimental bounds on the NCR
\cite{E,G,H,I}.  The confusion surrounding the r\^ole of the box may
be amplified by arguments supporting  the plausible scenario according
to which parts of the ``pure'' box, which could  be obtained after the
integration  over virtual   momenta,   may be  incorporated   into the
definition of the NCR without violating any of the physical principles
in  any obvious  way.  (Notice   that  the  PT  rearrangement of   the
amplitude  takes place   {\it before}  any  integrations over  virtual
momenta have been carried out).

In this  paper we demonstrate that the   exclusion of the  ``pure'' or
``effective'' $WW$ box from the definition of the NCR  is not a matter
of convention  but  is in  fact dynamically  realized in a  particular
physical  process, namely  when the  target-fermions are right-handedly
polarized, simply because in  that case the $WW$ box  is not  there to
begin   with.  In  particular we demonstrate    explicitly that the PT
construction  goes through unmodified   in  the absence  of  the  box,
because the cancellations previously enforced by  parts of the box are
now enforced by  the modification of the  couplings of the photon  and
the $Z$  to right-handed fermions.  In  particular,  at tree-level the
$\gamma   f_{\chic  R}  \bar{f}_{\chic   R}$   and   $Z   f_{\chic  R}
\bar{f}_{\chic R}$ couplings  allow  for a cancellation involving  the
two $s$-channels graphs, one mediated  by an off-shell photon, and one
by  an off-shell $Z$.  This cancellation  replaces the usual $s$-channel
-- $t$-channel cancellation     occuring in the  case   of unpolarized
fermions \cite{PP2,PRW}, since 
now there    is no $t$-channel graphs   involving  two
outgoing $W$.  The aforementioned tree-level cancellation is
in   fact responsible for     the  one-loop
cancellations we will demonstrate in detail.  The analysis carried out
settles  conclusively the    question of   whether   after the   gauge
cancellations have been implemented the pure  box (or some arbitrarily
chosen parts of  it)  should be part of   the definition  of  the form
factor giving       rise to the   NCR.     In  addition,  the explicit
demonstration of the gauge cancellations in two different gauge-fixing
schemes   ($R_{\xi}$   and   BFM)  shows  the   general  validity  and
self-consistency of the  entire  procedure, and, as a  by-product,  it
makes abundantly clear the fact that (in the latter  case) at no point
the  choice   $\xi^{\chic  W}_{\chic  Q}   =   1$  has   been a-priori
implemented.

At the end of this procedure,  we obtain a neutrino electromagnetic 
form-factor  which (a) is independent of the
GFP; (b) couples electromagnetically to the target;  (c) it is process
(target)-independent and can  therefore be considered  as an intrinsic
property of  the neutrino; (d)   the  effective one-loop  $\gamma  \nu
\bar{\nu}$ proper vertex from which it  originates satisfies 
a QED-like WI; (e)
it holds for  arbitrary values (time-like and  
space-like ) of the four-momentum transfer
(f) since the self-energies do not enter into its definition,  
the resulting NCR does not depend on the quark sector of the theory. 

The paper is organized  as follows: In section  2 we present a general
discussion on the PT. In Section 3 we demonstrate explicitly the gauge
cancellations in the context  of the $R_{\xi}$ gauges, for unpolarized
target  fermions; here the  $WW$ box appears  and its participation in
the  aforementioned    cancellations is   crucial.  In section    4 we
demonstrate how the crucial cancellations are  enforced in the case of
right-handedly  polarized target fermions, i.e. when  the  $WW$ box is
actually absent. In   section 5 we  present  in detail how  the  above
cancellations are enforced in the context of  the BFM. In section 6 we
show how the final  rearrangement of the physical amplitude  according
to the general  PT rules singles out  the final gauge-invariant answer
for the  individual effective self-energies,  vertices, and boxes; the
latter   quantities    coincide with   their   conventionally  defined
counterparts calculated in the Feynman gauge  of the BFM.  
In section 7 we present explicit results for the
 NCR extracted from the effective photon-neutrino vertex,
and the slope (at $q^2=0$)
of the effective charge defined through the 
PT $\gamma Z$  self-energy.   
Finally, in section 8 we present our conclusions.

\setcounter{equation}{0}
\section{General discussion}

In this section we outline the 
field-theoretical mechanism which enforces 
the massive gauge cancellations and allows for 
the construction of
GFP-independent and gauge-invariant boxes, vertices, and self-energies. 
The NCR will then be defined from the  one-loop 
$\gamma \nu \nu $ vertex so constructed.

The longitudinal momenta appearing in the $S$-matrix element of
$F\bar{F}\to \nu \bar{\nu}$ originate from the 
tree-level gauge-boson propagators and tri-linear gauge-boson vertices
appearing inside loops. In particular, in the $R_{\xi}$-scheme
the gauge-boson propagators have the general form  
\be
\Delta^{\mu\nu}_{\chic V} (k) = \Bigg [g^{\mu\nu}-
\frac{(1-\xi_{\chic V})k^{\mu}k^{\nu}}{ k^2 -\xi_{\chic V} 
M_{\chic V}^2}\Bigg ]D_{\chic V}(k)
\label{GenProp}
\ee
with
\be
D_{\chic V}(k)= (k^2 -M_{\chic V}^2)^{-1}
\label{Deno}
\ee
where $V=W,N$  with $N =  Z,\gamma$ and $M^2_{\gamma}=0$;  $k$ denotes
the virtual four-momentum circulating in the loop.  Clearly,  in the case of
$\Delta^{\mu\nu}_{\chic V} (k)$  the  longitudinal momenta are   those
proportional to $(1-\xi_{\chic V})$.
The longitudinal terms arising from the tri-linear vertex 
may be identified by splitting 
$\Gamma_{\alpha\mu\nu}(q,k,-k-q)$
appearing inside the one-loop diagrams (see Fig.1a and Fig.2 for the
definition of the momenta entering into the vertex) 
into two parts ($q$ denotes the physical four-momentum entering
into the vertex)\cite{TH}:
\bea
\Gamma_{\alpha\mu\nu} (q,k,-k-q) 
&=& (q-k)_{\nu} g_{\alpha\mu} + (2k+q)_{\alpha} g_{\mu\nu} 
- (2q+k)_{\mu} g_{\alpha\nu} \nonumber\\  
&=& \Gamma_{\alpha\mu\nu}^{\chic F} + \Gamma_{\alpha\mu\nu}^{\chic P} \, ,
\label{decomp}
\eea
with 
\bea
\Gamma_{\alpha\mu\nu}^{\chic F}&=& 
(2k+q)_{\alpha} g_{\mu\nu} + 2q_{\nu}g_{\alpha\mu} 
- 2q_{\mu}g_{\alpha\nu} \, , \nonumber\\
\Gamma_{\alpha\mu\nu}^{\chic P} &=&
 -(k+q)_{\nu} g_{\alpha\mu} - k_{\mu}g_{\alpha\nu} \, .  
\label{GFGP}
\eea
The
first term in $\Gamma_{\alpha\mu\nu}^{\chic F}$ is a convective
vertex describing the coupling of a vector boson to a scalar field, 
whereas the
other two terms originate from spin or magnetic moment. 
The above decomposition assigns a special r\^ole 
to the $q$-leg,
and allows $\Gamma_{\alpha\mu\nu}^{\chic F}$ 
to satisfy the Ward identity
\be 
q^{\alpha} \Gamma_{\alpha\mu\nu}^{\chic F}= 
 (k+q)^2 g_{\mu\nu} - k^2  g_{\mu\nu}\, ,
\label{WI2B}
\ee
All aforementioned longitudinal momenta 
originating from $\Delta^{\mu\nu}_V (k)$ 
and $\Gamma_{\alpha\mu\nu}^{\chic P}$ 
trigger the following 
WI when contracted with the appropriate $\gamma$ matrix
appearing in the various elementary vertices :
\bea
\not\! k  P_{\chic L} &=& (\not\! k + \not\! p ) P_{\chic L} 
- P_{\chic R} \not\! p \, ,
\nonumber\\
&=& S_{\chic F'}^{-1}(\not\! k + \not\! p ) P_{\chic L} - 
P_{\chic R} S_{\chic F}^{-1}(\not\! p) + m_{\chic F'} P_{\chic L}  
-  m_{\chic F} P_{\chic R}
\label{EWI}
\eea
where $P_{\chic R(L)} = [1  + (-) \gamma_5]/2$  is the  chirality projection
operator and   
$S_{\chic F}$ is the tree-level propagator of the fermion $F$;
$F'$ is the
isodoublet-partner of the external fermion $F$.
The result of this contraction is that 
the term in  Eq.(\ref{EWI}) proportional to  $S_{\chic F'}^{-1}$, i.e
the inverse of the internal fermion propagator      
gives rise to a self-energy-like term, 
whose coupling to the external fermions is proportional
to the vertex 
\be
\Gamma_{{\chic W} {\chic F} \bar{\chic F}}^{\mu} 
= -i\bigg(\frac{g_w }{2}\bigg) \gamma_\mu P_{\chic L}.
\ee
The effective vertex $\Gamma_{{\chic W} {\chic F} \bar{\chic F}}^{\mu}$ 
should not be confused with the usual $WFF'$ elementary vertex
involving two fermions $F$ and $F'$ of different charge and isospin.
The vertex  $\Gamma_{{\chic W} {\chic F} \bar{\chic F}}^{\mu}$ can be 
written as a linear combination of the two tree-level
vertices $\Gamma_{\gamma {\chic F} \bar{\chic F}}^{\mu}$ and
$\Gamma_{{\chic Z} {\chic F} \bar{\chic F}}^{\mu}$ given by
\bea
\Gamma_{\gamma {\chic F} \bar{\chic F}}^{\mu} &=& -ig_w s_w 
Q_{\chic F} \gamma^{\mu} \nonumber\\ 
\Gamma_{{\chic Z} {\chic F} \bar{\chic F}}^{\mu} &=& \, 
-i \bigg(\frac{g_w}{c_w}\bigg)\, \gamma^{\mu}\, 
[ (s^2_w Q_{\chic F} - T^{\chic F}_z) P_{\chic L} + s^2_w Q_{\chic F} 
P_{\chic R}]
\label{GenVer}
\eea
as follows:
\be
\Gamma_{{\chic W} {\chic F} \bar{\chic F}}^{\mu}  
= \bigg(\frac{s_w}{2 T^{\chic F}_z}\bigg)
\Gamma_{\gamma {\chic F} \bar{\chic F}}^{\mu} - 
\bigg(\frac{c_w}{2 T^{\chic F}_z}\bigg)\Gamma_{{\chic Z} {\chic F}
\bar{\chic F}}^{\mu}
\label{SepVer}
\ee
In the above formulas
$Q_{\chic F}$ is the electric charge of the fermion $F$, and
$T^{\chic F}_z$ its $z$-component of the weak isospin,
with $c_w = \sqrt{1 - s^2_w} = M_{\chic W}/M_{\chic Z}$, $e=g_w s_w$,

The identity presented in Eq.(\ref{SepVer})
allows to combine the propagator-like
parts  emerging from box-diagrams and vertex-diagrams
after the application of the WI in  Eq.(\ref{EWI})
with the conventional 
self-energy graphs $\Pi^{\gamma {\chic Z}}_{\mu\nu}$ and
$\Pi^{{\chic Z} {\chic Z}}_{\mu\nu}$ (Fig.~1), by judiciously multiplying by 
$D_{\chic N} (q) D_{\chic N}^{-1}(q)$, as shown in Fig.~2 -- Fig.~4 .
In fact, in order to see exactly against which parts
of the self-energies the aforementioned terms will
cancel, one must notice  \cite{NJW2} 
that every time there is 
longitudinal momentum in a vertex or a box diagram
triggering the WI in  Eq.(\ref{EWI}), the same 
longitudinal momentum inside a self-energy diagram
will be contracted with one of the elementary tri-linear
vertices, triggering a 
WI of the form
\bea
(k+q)^{\nu} \Gamma_{\alpha\mu\nu} (q,k,-k-q)
&=& (q^2 g_{\alpha\mu} -q_{\alpha}q_{\mu}) - 
(k^2 g_{\alpha\mu} - k_{\alpha}k_{\mu}) \nonumber \\ 
&=& \bigg ( D_{\chic N}^{-1}(q)  -  D_{\chic W}^{-1}(k) 
+ (M_{\chic N}^2-M_{\chic W}^2)\bigg ) g_{\alpha\mu} +  k_{\alpha}k_{\mu}
\label{WI3g}
\eea
The terms proportional to $D_{\chic N}^{-1}(q)$ appearing in  
Eq.(\ref{WI3g}) are precisely those which will cancel
against the aforementioned propagator-like pieces coming from
vertices and boxes. It is clear that the exploitation of
the elementary WI of Eq.(\ref{EWI}) and  Eq.(\ref{WI3g})  enable
the communication between Feynman diagrams which at first 
seem to be kinematically distinct. 

It turns out that the one-loop
cancellations described above descend in fact 
from a fundamental tree-level cancellation between
$s$- and $t-$ channel diagrams, the latter 
being a consequence of the underlying BRS symmetry
\cite{PP2,PRW}.
For the remainder of this paper we will restrict our attention to the 
case where the fermions $F$ are leptons, to be denoted by $f$; 
in that case $T_z^f = -\frac{1}{2}$ and for the neutrino 
$T_z^{\nu_e} = \frac{1}{2}$
and so the identity becomes
\bea
\Gamma_{{\chic W} {\chic f} \bar{\chic f}}^{\mu}  &=& -s_w 
\Gamma_{\gamma {\chic f} \bar{\chic f}}^{\mu}  
+ c_w \Gamma_{{\chic Z} {\chic f} \bar{\chic f}}^{\mu}\nonumber\\  
\Gamma_{{\chic W} \nu\bar{\nu}}^{\mu}  &=& - c_w 
\Gamma_{{\chic Z} \nu \bar{\nu}}^{\mu}
\label{abc}
\eea
To see how the aforementioned tree-level cancellation is
enforced, consider the process $f\bar{f} \to W^{+} W^{-}$, 
whose tree-level amplitude will be denoted by ${\cal T}_{\mu\nu} $.
Defining
\be
{\cal V}_{{\chic V} {\chic f} \bar{\chic f}}^{\mu}= 
\bar{v}_{\chic f} (p_2)\, \Gamma_{{\chic V} 
{\chic f} \bar{\chic f}}^{\mu}\  u_f(p_1) 
\label{VerVFF}
\ee
we have that ${\cal T}_{\mu\nu}$ is given by two $s$-channel
and one $t$-channel graphs, 
\be
{\cal T}_{\mu\nu} = {\cal T}_{s\, \mu\nu}^{(\gamma)} +
{\cal T}_{s\, \mu\nu}^{(\chic Z)} +
{\cal T}_{{t}\  {\mu\nu}}
\label{k4m}
\ee
where 
\bea
{\cal T}_{s\, \mu\nu}^{(\gamma)} &=& {\cal V}_{\gamma {\chic f}
\bar{\chic f}}^{\alpha}
  D_\gamma(q)\Gamma_{\alpha\mu\nu}^{\gamma {\chic W} {\chic W}} 
(q,k_1,k_2)\, ,\nonumber \\  
{\cal T}_{s\, \mu\nu}^{(\chic Z)}  &=& {\cal V}_{{\chic Z} 
{\chic f} \bar{\chic f}}^{\alpha}
  D_{\chic Z} (q)\Gamma_{\alpha\mu\nu}^{{\chic Z} {\chic W} {\chic W}} 
(q,k_1,k_2)\, ,\nonumber\\
{\cal T}^{(t)}_{{\mu\nu}} &=& -\frac{g_w^2}{2}\, \bar{v}_{\chic f} (p_2)\, 
\gamma^\mu P_{\chic L}\, 
  S_{\chic f'}(\not\! p_1-\not\! k_1)\,  \gamma^\nu P_{\chic L}\,  
u_{\chic f}(p_1)\, ,
\label{k3m}
\eea
with 
\bea
\Gamma_{\alpha\mu\nu}^{{\gamma} {\chic W} {\chic W}} (q,k_1,k_2) &=& -g_w s_w 
\Gamma_{\alpha\mu\nu} (q,k_1,k_2)
\nonumber \\  
\Gamma_{\alpha\mu\nu}^{ {\chic Z} {\chic W} {\chic W}} (q,k_1,k_2) &=& g_w c_w 
\Gamma_{\alpha\mu\nu} (q,k_1,k_2)
\label{k2m}
\eea
Here,   
$s= q^{2} = (p_1+p_2)^{2}=(k_1+k_2)^2$  is the   c.m.\  energy squared.

Let us now consider the case where ${\cal T}_{\mu\nu}$ is contracted
by a longitudinal momentum 
$k_{1}^{\mu}$ or $k_{2}^{\nu}$.
Then, 
the WI of Eq.(\ref{WI3g}) will operate at the two $s$-channel graphs,
whereas that of  Eq.(\ref{EWI}) at the  $t$-channel graph (Fig.~5), yielding 
\bea
k_{1 \mu} {\cal T}^{\mu\nu}_{s\, (\gamma)} &=& - s_w {\cal V}^{\nu}_
{\gamma {\chic f} \bar{\chic f}}
 + S^{(\gamma) \nu}
\, ,\nonumber \\  
k_{1 \mu} {\cal T}^{\mu\nu}_{s\, (\chic Z)}  &=& c_w 
{\cal V}^{\nu}_{{{\chic Z}  {\chic f} \bar{\chic f}}}
+ S^{(\chic Z) \nu} 
\, ,\nonumber\\
k_{1 \mu} {\cal T}_{(t)}^{\mu\nu} &=& {\cal V}^{\nu}_{{{\chic W} 
{\chic f} \bar{\chic f}}}
\label{k1m} 
\eea
with 
\bea
S^{({\chic Z})}_{\nu} &=& {\cal V}_{{\chic Z} {\chic f} 
\bar{\chic f}}^{\alpha}D_{\chic Z}(q) c_w \bigg(
(M_{\chic Z}^2-M_{\chic W}^2)g_{\alpha\nu}+ 
k_{2\alpha}k_{2\nu}\bigg)\nonumber\\
S^{(\gamma)}_{\nu} &=&  {\cal V}_{\gamma {\chic f} 
\bar{\chic f}}^{\alpha}D_{\chic Z}(q) s_w \bigg(
M_{\chic W}^2 g_{\alpha\nu}- k_{2\alpha}k_{2\nu}\bigg)
\eea

Adding by parts both sides of Eq.(\ref{k1m})
we see that a major cancellation takes place; in particular
the pieces containing the vertices ${\cal V}_{Vf\bar{f}}^{\alpha}$
cancel by virtue of the first identity in Eq.(\ref{abc}), and
one is left on the RHS with purely $s$-channel contribution, i.e 
\be
k_1^{\mu} {\cal T}_{\mu\nu} = S^{(\chic Z)}_{\nu} + S^{(\gamma)}_{\nu}  
\label{k6m} 
\ee

It is important to   recognize that the cancellation   described above
goes  through  even when the  initial  fermions  are right-handedly
polarized, although in that  case  there is no  $t$-channel graph 
(Fig.~6). In fact, 
the  elementary  vertices given in Eq.(\ref{GenVer}),
together   with  the  corresponding   ${\cal V}_{\gamma {\chic f}
\bar{\chic f}}^{\mu}$ and ${\cal   V}_{{\chic Z} {\chic f} \bar{\chic f}}
^{\mu}$    are   appropriately    modified.  In particular,
\bea
\Gamma^{\mu}_{\gamma {\chic f}_{\chic R} \bar{\chic f}_{\chic R}} &=& 
-ig_w s_w Q_{\chic f} \gamma^{\mu} \nonumber\\ 
\Gamma^{\mu}_{ {\chic Z} {\chic f}_{\chic R} \bar{\chic f}_{\chic R}} 
&=& \, -i \bigg(\frac{g_w}{c_w}\bigg) s^2_w Q_{\chic f} \, \gamma^{\mu}\, 
\label{GenVerPol}
\eea
and
\be
{\cal V}^{\mu}_{{\chic Z} {\chic f}_{\chic R} \bar{\chic f}_{\chic R}} = 
 \bigg(\frac{s_w}{c_w}\bigg) {\cal V}^{\mu}_{ \gamma {\chic f}_{\chic R} 
\bar{\chic f}_{\chic R}}
\label{xyz}
\ee
Clearly, in that case, from Eq.\ (\ref{abc}) follows immediately that 
\be
{\cal V}_{{\chic W} {\chic f}_{\chic R} \bar{\chic f}_{\chic R}}^{\mu} = 0 .
\label{Vwfr}
\ee
and Eq.\ (\ref{k6m}) is still satisfied.

\section{The unpolarized case in the $R_{\xi}$ gauge}

In this section we will demonstrate explicitly how the cancellation of
the GFP proceeds in  the  case when the  theory  is quantized  in  the
framework of the renormalizable $R_{\xi}$ gauge.   Of course, based on
general field-theoretical  principles,  one knows in advance  that the
entire amplitude  will be GFP-independent.  What we will show  in this
section is that in fact this cancellation  goes through without having
to     carry  out  any    of    the integrations  over   virtual  loop
momenta. Indeed, even though  individual graphs conspire non-trivially
in order for  the  cancellations to  take place, the  way this happens
kinematically  is  very distinct.   In  particular, all  GFP-dependent
parts are  self-energy like; their  identification and extraction from
box and vertex diagrams is  implemented exclusively through the use of
elementary WI.  Therefore, the remaining GFP-independent parts of each
diagram retain their original kinematic identity,  i.e.  one may still
identify box-, vertex-, and self-energy-like sub-amplitudes
\cite{Var}.

We will focus on the case where the target fermions are 
charged leptons, which we
denote by $f$;
the generalization to the case where the target fermions are quarks 
is straightforward. 
In what follows
we will
set  $\lambda_{\chic W} \equiv 1-\xi_{\chic W}$, and we will suppress 
a factor
$g_w^2 \int \frac{d^n k} {(2\pi)^n}$, where $n=4-\epsilon$
is the dimension of space-time
in dimensional regularization.
In addition we define
\bea
I_0 &=& \Bigg [\bigg(k^2 - M_{\chic W}^2\bigg) 
\bigg((k+q)^2- M_{\chic W}^2\bigg)\Bigg]^{-1}
\nonumber\\
I_1  &=& \Bigg [\bigg(k^2 -\xi_{\chic W} M_{\chic W}^2\bigg)
\bigg (k^2-M_{\chic W}^2 \bigg)\Bigg]^{-1}
\nonumber\\
I_2 &=& \Bigg [\bigg(k^2 - \xi_{\chic W} M_{\chic W}^2\bigg) 
\bigg((k+q)^2-\xi_{\chic W} M_{\chic W}^2\bigg)\Bigg]^{-1}
\nonumber\\
I_3 &=& \Bigg [\bigg(k^2 -\xi_{\chic W} M_{\chic W}^2\bigg)
\bigg(k^2 -M_{\chic W}^2\bigg) 
\bigg((k+q)^2-M_{\chic W}^2\bigg)\Bigg]^{-1}
\nonumber\\ 
I_4 &=& \Bigg [\bigg(k^2 -\xi_{\chic W} M_{\chic W}^2\bigg)
\bigg((k+q)^2-\xi_{\chic W} M_{\chic W}^2\bigg)\bigg(k^2 -M_{\chic W}^2\bigg) 
\bigg((k+q)^2-M_{\chic W}^2\bigg)\Bigg]^{-1}\nonumber\\ 
I_5  &=& \bigg(k^2 -\xi_{\chic W} M_{\chic W}^2\bigg)^{-1}
\label{DefIs}
\eea
and  
we will employ the algebraic identity
\be
\frac{1}{k^2 -\xi_{\chic W} M_{\chic W}^2} = \frac{1}{k^2 -M_{\chic W}^2}
-\frac{(1-\xi_{\chic W})M_{\chic W}^2}{ (k^2 -M_{\chic W}^2)
(k^2 -\xi_{\chic W} M_{\chic W}^2)} 
\label{AlgId}
\ee
in order to rearrange various expressions. In addition, 
we
will drop terms proportional to $q_{\mu}$ or $q_{\nu}$
because we assume conserved external currents 
(mass-less fermions)

Turning to individual diagrams,  we first 
consider the box graphs shown in Fig.~7. We have:
\bea
(n) &=& (n)_{\xi_{\chic W} = 1} +  
{\cal V}_{{\chic W} {\chic f} \bar{\chic f}}^{\mu}\Bigg( 
\lambda_{\chic W}^2 I_4 k_{\mu}k_{\nu}
-2\lambda_{\chic W} I_3 g_{\mu\nu} \Bigg){\cal V}_{{\chic W} 
\nu\bar{\nu}}^{\nu}
\nonumber\\  
&=& (n)_{\xi_{\chic W} =1} + (n)_{\gamma {\chic Z}} 
+ (n)_{{\chic Z} {\chic Z}}
\label{Defn}
\eea
with
\bea
(n)_{\gamma {\chic Z}} &=&  
{\cal V}_{\gamma {\chic f} \bar{\chic f}}^{\mu}D_{\gamma} (q)
\Bigg[ s_w c_w D_{\gamma}^{-1}(q)
 \Bigg( \lambda_{\chic W}^2 I_4 k_{\mu}k_{\nu}
-2\lambda_{\chic W} I_3 g_{\mu\nu}\Bigg)D_{\chic Z}^{-1}(q)\Bigg ]
D_{\chic Z}(q){\cal V}_{{\chic Z} \nu\bar{\nu}}^{\nu} \nonumber\\
(n)_{{\chic Z} {\chic Z}} &=& - 
{\cal V}_{{\chic Z} {\chic f} \bar{\chic f}}^{\mu}D_{\chic Z}(q)
\Bigg[ c_w^2 D_{\chic Z}^{-1}(q)
 \Bigg( \lambda_{\chic W}^2 I_4 k_{\mu}k_{\nu}
-2\lambda_{\chic W} I_3 g_{\mu\nu} \Bigg )D_{\chic Z}^{-1}(q)\Bigg] 
D_{\chic Z}(q){\cal V}_{{\chic Z} \nu\bar{\nu}}^{\nu}
\label{nGZnZZ}
\eea
Notice that $(n)_{\gamma {\chic Z}}$ and $(n)_{{\chic Z} {\chic Z}}$ 
are propagator-like quantities

For the vertex graphs $(o)$, $(p)$, and $(q)$ we have
(there is a relative minus sign from the difference in the 
$ZWW$ and $\gamma WW$ vertices, as given in  Eq.(\ref{k2m})):
\bea
(o) &=& (o)_{\xi_{\chic W} =1} + 
{\cal V}_{\gamma {\chic f} \bar{\chic f}}^{\mu}D_{\gamma} (q) 
s_w \Bigg(\lambda_{\chic W}^2 I_4 q^2 k_{\mu}k_{\nu} -2 \lambda_{\chic W} 
I_3 \bigg [q^2g_{\mu\nu} - (k+q)^2g_{\mu\nu} + k_{\mu}k_{\nu}\bigg ] \Bigg)
{\cal V}_{{\chic W} \nu\bar{\nu}}^{\nu}\nonumber\\ 
(p) &=& (p)_{\xi_{\chic W} =1} -
{\cal V}_{{\chic Z} {\chic f} \bar{\chic f}}^{\mu}D_{\chic Z} (q) 
c_w \Bigg(\lambda_{\chic W}^2 I_4 q^2 k_{\mu}k_{\nu} -2 \lambda_{\chic W} 
I_3 [q^2g_{\mu\nu} - (k+q)^2g_{\mu\nu} + k_{\mu}k_{\nu}] \Bigg)
{\cal V}_{{\chic W} \nu\bar{\nu}}^{\nu}\nonumber\\ 
(q) &=& (q)_{\xi_{\chic W} =1} -
{\cal V}_{{\chic W} {\chic f} \bar{\chic f}}^{\mu}
c_w \Bigg(\lambda_{\chic W}^2 I_4 q^2 k_{\mu}k_{\nu} -2 \lambda_{\chic W} 
I_3 \bigg [q^2g_{\mu\nu} - (k+q)^2g_{\mu\nu} + k_{\mu}k_{\nu}\bigg ] \Bigg)
D_{\chic Z} (q) {\cal V}_{{\chic Z} \nu \bar{\nu}}^{\mu}\nonumber\\
&& {}
\label{opq}
\eea
Notice that the parts depending on $\xi_{\chic W}$ are again propagator-like.
To further exploit this fact, we use the identities of Eq.(\ref{abc}) to 
rewrite the vertices ${\cal V}_{{\chic W} {\chic f} \bar{\chic f}}^{\mu}$ and  
${\cal V}_{{\chic W} \nu\bar{\nu}}^{\nu}$ in terms of the fundamental ones, 
thus obtaining
\bea
(o) &=& (o)_{\xi_{\chic W} =1} + (o)_{\gamma {\chic Z}} \nonumber\\
(p) &=& (p)_{\xi_{\chic W} =1} + (p)_{{\chic Z} {\chic Z}}  \nonumber\\
(q) &=& (q)_{\xi_{\chic W} =1} + (q)_{\gamma {\chic Z}} 
+ (q)_{{\chic Z} {\chic Z}} \nonumber\\
\label{Ropq}
\eea
with
\bea
(o)_{\gamma {\chic Z}} &=& -{\cal V}_{\gamma {\chic f} 
\bar{\chic f}}^{\mu}D_{\gamma} (q)
\Bigg[s_w c_w \Bigg(\lambda_{\chic W}^2 D_{\gamma}^{-1}(q) 
I_4 k_{\mu}k_{\nu} - 2 \lambda_{\chic W} I_3 \bigg [D_{\gamma}^{-1}(q) 
g_{\mu\nu} - (k+q)^2 g_{\mu\nu} + k_{\mu}k_{\nu}\bigg ] \Bigg) \nonumber\\
&& D_{\chic Z}^{-1}(q)\Bigg]
\nonumber\\
&& D_{\chic Z}(q){\cal V}_{{\chic Z} \nu\bar{\nu}}^{\nu} \nonumber\\
(p)_{{\chic Z} {\chic Z}} &=& {\cal V}_{{\chic Z} {\chic f} \bar{\chic f}}
^{\mu}D_{\chic Z} (q)\Bigg[c_w^2 \Bigg(
\lambda_{\chic W}^2 \bigg\{D_{\chic Z}^{-1}(q) + M_{\chic Z}^2 \bigg \} 
I_4 k_{\mu}k_{\nu} - 2 \lambda_{\chic W} I_3 
\bigg [D_{\chic Z}^{-1}(q)g_{\mu\nu} + M_{\chic Z}^2 g_{\mu\nu}\nonumber\\ 
&& -(k+q)^2g_{\mu\nu} + k_{\mu}k_{\nu}\bigg ] \Bigg)
D_{\chic Z}^{-1}(q)\Bigg] D_{\chic Z}(q){\cal V}_
{{\chic Z} \nu\bar{\nu}}^{\nu} \nonumber\\
\nonumber\\
(q)_{\gamma {\chic Z}} &=& -{\cal V}_{\gamma {\chic f} \bar{\chic f}}
^{\mu}D_{\gamma} (q) \Bigg[s_w c_w D_{\gamma}^{-1}(q) \Bigg(
\lambda_{\chic W}^2 \bigg\{D_{\chic Z}^{-1}(q) 
+ M_{\chic Z}^2 \bigg \} I_4 k_{\mu}k_{\nu} 
 -2 \lambda_{\chic W} I_3 \bigg [D_{\chic Z}^{-1}(q)g_{\mu\nu} 
+M_{\chic Z}^2 g_{\mu\nu}\nonumber\\ 
&& -(k+q)^2g_{\mu\nu} + k_{\mu}k_{\nu}\bigg ]
\Bigg) \Bigg]
D_{\chic Z} (q) {\cal V}_{{\chic Z} \nu \bar{\nu}}^{\nu}
\nonumber\\
(q)_{{\chic Z} {\chic Z}} &=& 
{\cal V}_{{\chic Z} {\chic f} \bar{\chic f}}^{\mu} D_{\chic Z}(q)\Bigg[c_w^2
D_{\chic Z}^{-1}\Bigg(
\lambda_{\chic W}^2 \bigg\{D_{\chic Z}^{-1}(q) 
+ M_{\chic Z}^2 \bigg \}I_4 k_{\mu}k_{\nu} 
 -2 \lambda_{\chic W} I_3 \bigg [D_{\chic Z}^{-1}(q)g_{\mu\nu} 
+M_{\chic Z}^2 g_{\mu\nu}\nonumber\\ 
&& -(k+q)^2g_{\mu\nu} + k_{\mu}k_{\nu}\bigg ]
\Bigg) \Bigg]
D_{\chic Z} (q) {\cal V}_{{\chic Z} \nu \bar{\nu}}^{\nu}
\label{Eopqq}
\eea
where we have judiciously rewritten $q^2 = D_{\gamma}^{-1}(q)$
or $q^2 = D_{\chic Z}^{-1}(q) + M_{\chic Z}^2$. Notice that 
$(p)_{{\chic Z} {\chic Z}} =(q)_{{\chic Z} {\chic Z}}$.

In addition, the remaining Abelian-like  vertex graphs 
together with the external fermion wave function 
corrections give:
\bea
(s)+(t) &=& [(s)+(t)]_{\xi_{\chic W} =1} + [(s)+(t)]_{{\chic Z} {\chic Z}} 
\nonumber\\
(r) &=&  (r)_{\xi_{\chic W} =1} + (r)_{\gamma {\chic Z}} \nonumber\\
 (u)+(v) &=&  ((u)+(v))_{\xi_{\chic W} =1} + ((u)+(v))_{{\chic Z} {\chic Z}}
+ [(u)+(v)]_{\gamma {\chic Z}} 
\label{struv}
\eea
with
\bea
[(s)+(t)]_{{\chic Z} {\chic Z}} &=& 
- {\cal V}_{{\chic Z} {\chic f} \bar{\chic f}}^{\mu}D_{\chic Z} (q) 
\Bigg[c_w^2 \lambda_{\chic W} D_{\chic Z}^{-1}(q) I_1 g_{\mu\nu}\Bigg] 
D_{\chic Z} (q) {\cal V}_{{\chic Z} \nu \bar{\nu}}^{\nu} \nonumber\\
(r)_{\gamma {\chic Z}}  &=& {\cal V}_{ \gamma {\chic f} \bar{\chic f}}^
{\mu}D_{\gamma} (q) \Bigg[s_w c_w \lambda_{\chic W} 
D_{\chic Z}^{-1}(q) I_1 g_{\mu\nu}\Bigg]
D_{\chic Z} (q) {\cal V}_{{\chic Z} \nu \bar{\nu}}^{\nu} \nonumber\\
((u)+(v))_{{\chic Z} {\chic Z}} &=& - {\cal V}_{{\chic Z} {\chic f} 
\bar{\chic f}}^{\mu}D_{\chic Z} (q) 
\Bigg[c_w^2 \lambda_{\chic W} D_{\chic Z}^{-1}(q) I_1 g_{\mu\nu}\Bigg] 
D_{\chic Z} (q) {\cal V}_{{\chic Z} \nu \bar{\nu}}^{\nu} \nonumber\\
 ((u)+(v))_{\gamma {\chic Z}} &=& {\cal V}_{ \gamma {\chic f} 
\bar{\chic f}}^{\mu}D_{\gamma} (q) 
\Bigg[s_w c_w  \lambda_{\chic W} D_{\gamma}^{-1}(q) I_1 g_{\mu\nu}\Bigg] 
D_{\chic Z} (q) {\cal V}_{{\chic Z} \nu \bar{\nu}}^{\nu}
\label{Estruv}
\eea
where again all $\xi_{\chic W}$ dependent contributions are 
effectively propagator-like. Notice that 
$[(s)+(t)]_{{\chic Z} {\chic Z}} = ((u)+(v))_{{\chic Z} {\chic Z}}$.

Next continue with the conventional self-energy diagrams.
We will focus first on (a) since a big part of its gauge dependent
part cancels the entire gauge dependence coming from the vertices
and the boxes, given above.
Straightforward calculation, using only the elementary Ward identities
triggered by the action of the longitudinal parts of the
internal $W$ propagators on the three-boson vertices yields:  
\bea
(a)_{{\chic Z} {\chic Z}} &=& (a)_{{\chic Z} {\chic Z}}^{(\xi_{\chic W}=1)} 
+ L_{{\chic Z} {\chic Z}}(\xi_{\chic W}) \nonumber\\ 
&& - {\cal V}_{{\chic Z} {\chic f} \bar{\chic f}}^{\mu} D_{\chic Z}(q)
\Bigg[ c_w^2 \Bigg (\lambda_{\chic W}^2 q^4 I_4 k_{\mu}k_{\nu}
-2\lambda_{\chic W} I_3 \bigg [q^4 g_{\mu\nu} + 2 q^2 k_{\mu}k_{\nu}    
-2q^2 (k+q)^2 g_{\mu\nu}\bigg]\Bigg)\Bigg]
D_{\chic Z} (q) {\cal V}_{{\chic Z} \nu \bar{\nu}}^{\nu}
\nonumber\\ 
(a)_{\gamma {\chic Z}} &=& (a)_{\gamma {\chic Z}}^{(\xi_{\chic W}=1)} 
+ L_{\gamma {\chic Z}}(\xi_{\chic W}) 
\nonumber\\ 
&& + {\cal V}_{ \gamma {\chic f} \bar{\chic f}}^{\mu} D_{\gamma}(q)
\Bigg[ s_w c_w \Bigg(\lambda_{\chic W}^2 q^4 I_4 k_{\mu}k_{\nu}
- 2\lambda_{\chic W} I_3 \bigg [q^4 g_{\mu\nu} + 2 q^2 k_{\mu}k_{\nu}    
-2q^2 (k+q)^2 g_{\mu\nu} \bigg ]\Bigg)\Bigg]
D_{\chic Z} (q) {\cal V}_{{\chic Z} \nu \bar{\nu}}^{\nu}\nonumber\\ 
&& {}
\label{aZZaGZ}
\eea
with
\bea
L_{{\chic Z} {\chic Z}}(\xi_{\chic W}) &=& 
{\cal V}_{{\chic Z} {\chic f} \bar{\chic f}}^{\mu} D_{\chic Z}(q)
\Bigg[2 c_w^2 \lambda_{\chic W} I_1 \Bigg( (k+q)^4 g_{\mu\nu} - 
(k+q)^2 k_{\mu}k_{\nu}\Bigg)\Bigg]D_{\chic Z}(q) 
{\cal V}_{{\chic Z} \nu \bar{\nu}}^{\nu}\nonumber\\ 
L_{\gamma {\chic Z}}(\xi_{\chic W}) &=&
-{\cal V}_{ \gamma {\chic f} \bar{\chic f}}^{\mu} D_{\gamma}(q)
\Bigg[2 s_w c_w \lambda_{\chic W} I_1 \Bigg( (k+q)^4 g_{\mu\nu} - 
(k+q)^2 k_{\mu}k_{\nu}\Bigg)\Bigg]D_{\chic Z} (q) 
{\cal V}_{{\chic Z} \nu \bar{\nu}}^{\nu}
\label{LZZLGZ}
\eea
Notice the appearance of $I_1$; this term has no $q$ dependence .

Next we will rewrite the expressions containing an explicit $q^2$
in terms of $D_{\chic Z}^{-1}(q)$ and $D_{\gamma}^{-1}(q)$:
\bea
(a)_{{\chic Z} {\chic Z}} &=& (a)_{{\chic Z} {\chic Z}}^{(\xi_{\chic W}=1)} 
+ L_{{\chic Z} {\chic Z}}(\xi_{\chic W}) 
+ R_{{\chic Z} {\chic Z}}(\xi_{\chic W}) \nonumber\\ 
&& -{\cal V}_{{\chic Z} {\chic f} \bar{\chic f}}^{\mu} D_{\chic Z}(q)
\Bigg[c_w^2 \Bigg (\lambda_{\chic W}^2 I_4 \bigg \{ D_{\chic Z}^{-2}(q)
+ 2 M_{\chic Z}^2 D_{\chic Z}^{-1}(q) \bigg \} k_{\mu}k_{\nu}
\nonumber\\
&& - 2 \lambda_{\chic W} I_3 \bigg [D_{\chic Z}^{-2}(q) g_{\mu\nu}
+ 2 D_{\chic Z}^{-1}(q) \bigg (k_{\mu}k_{\nu} + M_{\chic Z}^2 g_{\mu\nu} -  
(k+q)^2 g_{\mu\nu}\bigg) \bigg]\Bigg)\Bigg]
D_{\chic Z} (q) {\cal V}_{{\chic Z} \nu \bar{\nu}}^{\nu}
\nonumber\\ 
(a)_{\gamma {\chic Z}} &=& (a)_{\gamma {\chic Z}}^{(\xi_{\chic W}=1)} 
+ L_{\gamma {\chic Z}}(\xi_{\chic W}) + R_{\gamma {\chic Z}}(\xi_{\chic W})
\nonumber\\ 
&& + {\cal V}_{ \gamma {\chic f} \bar{\chic f}}^{\mu} D_{\gamma}(q)
\Bigg[s_w c_w \Bigg(
\lambda_{\chic W}^2 I_4 
\bigg \{  D_{\chic Z}^{-1}(q)D_{\gamma}^{-1}(q)
+ M_{\chic Z}^2 D_{\gamma}^{-1}(q)\bigg \} k_{\mu}k_{\nu} 
 - 2\lambda_{\chic W} I_3 
\bigg[ D_{\chic Z}^{-1}(q)D_{\gamma}^{-1}(q)g_{\mu\nu} \nonumber\\ 
&& + D_{\gamma}^{-1}(q) 
\bigg \{ k_{\mu}k_{\nu} + M_{\chic Z}^2 g_{\mu\nu} 
- (k+q)^2 g_{\mu\nu}\bigg \} 
+ D_{\chic Z}^{-1}(q) \bigg \{ k_{\mu}k_{\nu} - (k+q)^2 g_{\mu\nu}\bigg \} 
\bigg]\Bigg)\Bigg]
D_{\chic Z} (q) {\cal V}_{{\chic Z} \nu \bar{\nu}}^{\nu} \nonumber\\
&& {}
\label{RaZZ}
\eea
with
\bea
R_{{\chic Z} {\chic Z}}(\xi_{\chic W}) &=& 
-{\cal V}_{{\chic Z} {\chic f} \bar{\chic f}}^{\mu} D_{\chic Z}(q) 
\Bigg[c_w^2 \Bigg( \lambda_{\chic W}^2 I_4 M_{\chic Z}^4 k_{\mu}k_{\nu}
-2 \lambda_{\chic W} I_3 
\bigg \{ 2  M_{\chic Z}^2  \bigg(k_{\mu}k_{\nu} -(k+q)^2 g_{\mu\nu}\bigg)
+ M_{\chic Z}^4 g_{\mu\nu}\bigg \} 
\Bigg)\Bigg]
D_{\chic Z} (q) {\cal V}_{{\chic Z} \nu \bar{\nu}}^{\nu} \nonumber\\
R_{\gamma {\chic Z}}(\xi_{\chic W}) &=& 
-{\cal V}_{\gamma {\chic f} \bar{\chic f}}^{\mu} D_{\gamma}(q) 
\Bigg[2 s_w c_w \lambda_{\chic W} I_3 
M_{\chic Z}^2  \bigg(k_{\mu}k_{\nu} -(k+q)^2 g_{\mu\nu}\bigg)\Bigg]
D_{\chic Z} (q) {\cal V}_{{\chic Z} \nu \bar{\nu}}^{\nu} 
\label{RZZRGZ}
\eea

Adding the contributions from the diagrams 
$(n)$,$(o)$,$(p)$,$(q)$, $(a)_{{\chic Z} {\chic Z}}$, and 
$(a)_{\gamma {\chic Z}}$ a massive cancellation takes place, and we obtain:
\bea
(n)+(o)+(p)+(q)+(a)_{{\chic Z} {\chic Z}}+(a)_{\gamma {\chic Z}} &=&
(n)^{(\xi_{\chic W}=1)}+(o)^{(\xi_{\chic W}=1)}+ (p)^{(\xi_{\chic W}=1)}+ 
(q)^{(\xi_{\chic W}=1)} \nonumber\\ 
&& +(a)_{{\chic Z} {\chic Z}}^{(\xi_{\chic W}=1)} 
+ (a)_{\gamma {\chic Z}}^{(\xi_{\chic W}=1)}
+ R_{{\chic Z} {\chic Z}}(\xi_{\chic W}) 
+ L_{{\chic Z} {\chic Z}}(\xi_{\chic W})  
\nonumber\\
&& + R_{\gamma {\chic Z}}(\xi_{\chic W}) 
+ L_{\gamma {\chic Z}}(\xi_{\chic W})
\label{nopq}
\eea

Thus, at this point all  
$\xi_{\chic W}$-dependent pieces are contained in the sum of
$R_{{\chic Z} {\chic Z}}(\xi_{\chic W}) 
+ L_{{\chic Z} {\chic Z}}(\xi_{\chic W})$ and 
$R_{\gamma {\chic Z}}(\xi_{\chic W}) 
+ L_{\gamma {\chic Z}}(\xi_{\chic W})$;
we will next show how these terms 
cancel exactly against the terms given in Eq.(\ref{Estruv}))
and
the gauge-dependent terms of the remaining 
self-energy diagrams.

Indeed, straightforward algebra allows one to recast the sums
$R_{{\chic Z} {\chic Z}}(\xi_{\chic W}) 
+ L_{{\chic Z} {\chic Z}}(\xi_{\chic W})$ 
and $R_{\gamma {\chic Z}}(\xi_{\chic W}) 
+ L_{\gamma {\chic Z}}(\xi_{\chic W})$ in the following form:
\bea
R_{{\chic Z} {\chic Z}}(\xi_{\chic W}) 
+ L_{{\chic Z} {\chic Z}}(\xi_{\chic W})  &=& 
 -{\cal V}_{{\chic Z} {\chic f} \bar{\chic f}}^{\mu} D_{\chic Z}(q) 
\Bigg[c_w^2 \Bigg( \lambda_{\chic W}^2 I_4 M_{\chic Z}^4 k_{\mu}k_{\nu}
-2 \lambda_{\chic W}  \Bigg \{
 (k^2 g_{\mu\nu} -k_{\mu}k_{\nu})I_1 \nonumber\\ 
&& + (2M_{\chic Z}^2-M_{\chic W}^2) I_3 k_{\mu}k_{\nu} 
+ (M_{\chic Z}^2- M_{\chic W}^2)^2 I_3 g_{\mu\nu}
+ D_{\chic Z}^{-1}(q) I_1 g_{\mu\nu} \nonumber\\ 
&& + (M_{\chic W}^2- M_{\chic Z}^2)I_1 g_{\mu\nu} 
\Bigg \} 
\Bigg)\Bigg]
D_{\chic Z} (q) {\cal V}_{{\chic Z} \nu \bar{\nu}}^{\nu}
\label{RZZLZZ}
\eea
and
\bea
R_{\gamma {\chic Z}}(\xi_{\chic W}) + L_{\gamma {\chic Z}}(\xi_{\chic W}) &=& 
 -{\cal V}_{\gamma {\chic f} \bar{\chic f}}^{\mu} D_{\gamma}(q) 
\Bigg[2s_w c_w \lambda_{\chic W}  \Bigg \{
 (k^2 g_{\mu\nu} -k_{\mu}k_{\nu})I_1 
+ (M_{\chic Z}^2-M_{\chic W}^2) I_3 k_{\mu}k_{\nu} \nonumber\\ 
&&+ M_{\chic W}^2 (M_{\chic W}^2- M_{\chic Z}^2) I_3 g_{\mu\nu}
+ \frac{1}{2} D_{\chic Z}^{-1}(q) I_1 g_{\mu\nu} 
+ \frac{1}{2} D_{\gamma}^{-1}(q) I_1 g_{\mu\nu}
\nonumber\\ 
&& + \bigg(M_{\chic W}^2- \frac{1}{2} M_{\chic Z}^2\bigg)I_1 g_{\mu\nu} 
\Bigg \} \Bigg]
D_{\chic Z} (q) {\cal V}_{{\chic Z} \nu \bar{\nu}}^{\nu}
\label{RGZLGZ}
\eea
One notices immediately that the
terms proportional to $D_{\chic Z}^{-1}(q)$ and 
$D_{\gamma}^{-1}(q)$ in Eq.(\ref{RZZLZZ})) and Eq.(\ref{RGZLGZ}))
cancel exactly against the terms
in Eq.(\ref{Estruv})). 
From the rest of the self-energy diagrams, with the 
aid of the identity in  Eq.(\ref{AlgId})), we obtain  
\bea
(b)_{{\chic Z} {\chic Z}} + (c)_{{\chic Z} {\chic Z}} &=& 
(b)_{{\chic Z} {\chic Z}}^{(\xi_{\chic W}=1)} 
+ (c)_{{\chic Z} {\chic Z}}^{(\xi_{\chic W}=1)}
\nonumber\\ 
&& +{\cal V}_{{\chic Z} {\chic f} \bar{\chic f}}^{\mu} D_{\chic Z}(q)
\Bigg[2s_w^{4}M_{\chic Z}^{2} \Bigg(\lambda_{\chic W}^2 M_{\chic W}^{2} I_4  
k_{\mu}k_{\nu} - \lambda_{\chic W} I_3 \bigg\{M_{\chic W}^{2}g_{\mu\nu}+
k_{\mu}k_{\nu}
\bigg\} \Bigg)\Bigg]
D_{\chic Z} (q) {\cal V}_{{\chic Z} \nu \bar{\nu}}^{\nu} 
\nonumber\\ 
(d)_{{\chic Z} {\chic Z}} + (e)_{{\chic Z} {\chic Z}} &=& 
(d)_{{\chic Z} {\chic Z}}^{(\xi_{\chic W}=1)} 
+ (e)_{{\chic Z} {\chic Z}}^{(\xi_{\chic W}=1)}
\nonumber\\ 
&& +{\cal V}_{{\chic Z} {\chic f} \bar{\chic f}}^{\mu} D_{\chic Z}(q)
\Bigg[2c_w^{4}M_{\chic Z}^{2} 
\Bigg(\lambda_{\chic W}^2 M_{\chic W}^{2} I_4 
- 2\lambda_{\chic W} I_3 \Bigg)k_{\mu}k_{\nu} \Bigg]
D_{\chic Z} (q) {\cal V}_{{\chic Z} \nu \bar{\nu}}^{\nu} 
\nonumber\\ 
(f)_{{\chic Z} {\chic Z}} &=& (f)_{{\chic Z} {\chic Z}}^{(\xi_{\chic W}=1)} 
- {\cal V}_{{\chic Z} {\chic f} \bar{\chic f}}^{\mu} D_{\chic Z}(q)
\Bigg[(c_w^{2}-s_w^{2})^{2} M_{\chic Z}^{2} 
\Bigg(\lambda_{\chic W}^2 M_{\chic W}^{2} I_4 
- 2\lambda_{\chic W} I_3 \Bigg)k_{\mu}k_{\nu}
\Bigg] D_{\chic Z} (q) {\cal V}_{{\chic Z} \nu \bar{\nu}}^{\nu} 
\nonumber\\ 
(g)_{{\chic Z} {\chic Z}} &=& (g)_{{\chic Z} {\chic Z}}^{(\xi_{\chic W}=1)}
- {\cal V}_{{\chic Z} {\chic f} \bar{\chic f}}^{\mu} D_{\chic Z}(q) 
\Bigg[2 \lambda_{\chic W} c_w^2 (k^2 g_{\mu\nu} -k_{\mu}k_{\nu})I_1
\Bigg] D_{\chic Z} (q) {\cal V}_{{\chic Z} \nu \bar{\nu}}^{\nu} 
\nonumber\\
(h)_{{\chic Z} {\chic Z}} &=& (h)_{{\chic Z} {\chic Z}}^{(\xi_{\chic W}=1)}
-{\cal V}_{{\chic Z} {\chic f} \bar{\chic f}}^{\mu} D_{\chic Z}(q) \Bigg[
\frac{1}{2} \lambda_{\chic W} (c_w^2 - s_w^2)^2 M_{\chic Z}^{2} I_1 g_{\mu\nu}
\Bigg] D_{\chic Z} (q) {\cal V}_{{\chic Z} \nu \bar{\nu}}^{\nu} 
\nonumber\\
(i)_{{\chic Z} {\chic Z}} &=& (i)_{{\chic Z} {\chic Z}}^{(\xi_{\chic W}=1)} 
- {\cal V}_{{\chic Z} {\chic f} \bar{\chic f}}^{\mu} D_{\chic Z}(q) \Bigg[
\bigg (\frac{ M_{\chic Z}^{2}}{M_{\chic H}^{2}}\bigg)\lambda_{\chic W} 
k^2 I_1 g_{\mu\nu}
\Bigg] D_{\chic Z} (q) {\cal V}_{{\chic Z} \nu \bar{\nu}}^{\nu} 
\nonumber\\
(j)_{{\chic Z} {\chic Z}} &=& (j)_{{\chic Z} {\chic Z}}^{(\xi_{\chic W}=1)} + 
{\cal V}_{{\chic Z} {\chic f} \bar{\chic f}}^{\mu} D_{\chic Z}(q) \Bigg[
\bigg (\frac{ M_{\chic Z}^{2}}{M_{\chic H}^{2}} \bigg) \lambda_{\chic W} 
k^2 I_1 g_{\mu\nu} \Bigg] D_{\chic Z} (q) 
{\cal V}_{{\chic Z} \nu \bar{\nu}}^{\nu} 
\nonumber\\
(k)_{{\chic Z} {\chic Z}} &=& (k)_{{\chic Z} {\chic Z}}^{(\xi_{\chic W}=1)} 
+{\cal V}_{{\chic Z} {\chic f} \bar{\chic f}}^{\mu} D_{\chic Z}(q) \Bigg[
\frac{1}{2} \lambda_{\chic W} M_{\chic Z}^{2} I_1 g_{\mu\nu}
\Bigg] D_{\chic Z} (q) {\cal V}_{{\chic Z} \nu \bar{\nu}}^{\nu} 
\label{fghi}
\eea
and
\bea
(b)_{\gamma {\chic Z}} + (c)_{\gamma {\chic Z}} &=& 
(b)_{\gamma {\chic Z}}^{(\xi_{\chic W}=1)} 
+ (c)_{\gamma {\chic Z}}^{(\xi_{\chic W}=1)}
\nonumber\\ 
&& +{\cal V}_{ \gamma {\chic f} \bar{\chic f}}^{\mu} D_{\chic Z}(q)
\Bigg[2c_w s_w^{3} M_{\chic Z}^{2} \Bigg(\lambda_{\chic W}^2 
M_{\chic W}^{2} I_4  k_{\mu}k_{\nu} 
- \lambda_{\chic W} I_3 \bigg\{M_{\chic W}^{2}g_{\mu\nu} +
k_{\mu}k_{\nu} \bigg\} \Bigg)\Bigg]
D_{\chic Z} (q) {\cal V}_{\gamma \nu \bar{\nu}}^{\nu} 
\nonumber\\ 
(d)_{\gamma {\chic Z}} + (e)_{\gamma {\chic Z}} &=& 
(d)_{\gamma {\chic Z}}^{(\xi_{\chic W}=1)} 
+ (e)_{\gamma {\chic Z}}^{(\xi_{\chic W}=1)}
\nonumber\\ 
&& + {\cal V}_{\gamma {\chic f} \bar{\chic f}}^{\mu} D_{\chic Z}(q)
\Bigg[2 s_w c_w^{3} M_{\chic Z}^{2} 
\Bigg(\lambda_{\chic W}^2 M_{\chic W}^{2} 
I_4 -2\lambda_{\chic W} I_3 \Bigg)k_{\mu}k_{\nu} \Bigg]
D_{\chic Z} (q) {\cal V}_{{\chic Z} \nu \bar{\nu}}^{\nu} 
\nonumber\\ 
(f)_{\gamma {\chic Z}} &=& (f)_{\gamma {\chic Z}}^{(\xi_{\chic W}=1)} 
+ {\cal V}_{\gamma {\chic f} \bar{\chic f}}^{\mu} D_{\chic Z}(q)
\Bigg[2 c_w s_w (c_w^{2}-s_w^{2}) M_{\chic Z}^{2} 
\Bigg(\lambda_{\chic W}^2 M_{\chic W}^{2} 
I_4 -2\lambda_{\chic W} I_3 \Bigg)k_{\mu}k_{\nu} \Bigg]
D_{\chic Z} (q) {\cal V}_{{\chic Z} \nu \bar{\nu}}^{\nu} 
\nonumber\\ 
(g)_{\gamma {\chic Z}} &=& (g)_{\gamma {\chic Z}}^{(\xi_{\chic W}=1)}
+ {\cal V}_{\gamma {\chic f} \bar{\chic f}}^{\mu} 
D_{\gamma}(q) \Bigg[2 \lambda_{\chic W} s_w c_w (k^2 g_{\mu\nu} - 
k_{\mu}k_{\nu})I_1 \Bigg] D_{\chic Z} (q) 
{\cal V}_{{\chic Z} \nu \bar{\nu}}^{\nu} 
\nonumber\\
(h)_{\gamma {\chic Z}} &=& (h)_{\gamma {\chic Z}}^{(\xi_{\chic W}=1)}
+ {\cal V}_{\gamma {\chic f} \bar{\chic f}}^{\mu} D_{\gamma}(q) \Bigg[
2 c_w s_w \lambda_{\chic W} \bigg(M_{\chic W}^2 
- \frac{1}{2} M_{\chic Z}^2 \bigg) I_1 g_{\mu\nu}
\Bigg] D_{\chic Z} (q) {\cal V}_{{\chic Z} \nu \bar{\nu}}^{\nu} 
\label{fGZgGZ}
\eea
Using that $2s_w^{4}+2c_w^{4} - (c_w^{2}-s_w^{2})^{2} =1$
and that $(c_w^{2}-s_w^{2})^{2} = 1-4s_w^{2}c_w^{2}$ one can easily verify
that all remaining $\xi_{\chic W}$-dependent terms cancel.
Notice in particular that the terms proportional to 
$\lambda_{\chic W}^2$ in $(b)_{\gamma {\chic Z}} 
+ (c)_{\gamma {\chic Z}}$, $(d)_{\gamma {\chic Z}} 
+ (e)_{\gamma {\chic Z}}$, and $(f)_{\gamma {\chic Z}}$, cancel among 
 0: each other.

\setcounter{equation}{0}
\section{ The polarized case in the $R_{\xi}$ gauges}

We next turn to the case where the target fermions are
right-handedly polarized. As already mentioned, in that
case, due to the helicity mismatch, there is no coupling
of the $W$ to the fermions, and therefore there is no 
$WW$ box.  The purpose of this section is to demonstrate 
how the gauge cancellation go through in the
absence of the box. As we will see one arrives at
precisely the same expressions for the 
individual propagators and vertices 
which couple electromagnetically to the
target fermions as in the previous section. Thus,
even though the aforementioned vertices and 
propagators will be further rearranged 
following the PT algorithm, it is already clear
at this stage that the remaining box graphs simply cannot
enter in the (process-independent) definition of the NCR.

The mechanism which enforces the gauge cancellations in the absence of
the  box has  already been outlined  at  the end of    section 3.  The
modification  of  the   bare vertices    $\Gamma_{\gamma {\chic   f_R}
\bar{\chic f}_{\chic R}}^{\mu}$   and $\Gamma_{ {\chic  Z} {\chic f_R}
\bar{\chic f}_{\chic R} }^{\mu}$ given in Eq.(\ref{GenVerPol}) allows
the communication  between Feynman diagrams   whose only difference is
whether the  incoming gauge boson is a  photon or a $Z$  (for example
the diagrams  $(a)_{\gamma \chic Z}$   and $(a)_{\chic Z \chic Z}$  of
Fig.1,  respectively). In particular,  due to Eq.(\ref{GenVerPol})) the
sum of any  two such graphs  would cancel if  it was not for the  fact
that   the tree-level  propagators connecting   them   to the external
right-handed  fermions are different ($D_{\gamma}(q)$ for $(a)_{\gamma
\chic Z}$  , $D_{\chic Z}(q)$ for $(a)_{\chic  Z  \chic Z}$).  But the
gauge  dependent  pieces coming   from   these graphs are   precisely
proportional      to     inverse bare   
photon    and    $Z$   propagators,
($D_{\gamma}^{-1}(q)$ for $(a)_{\gamma \chic Z}$, $D_{\chic Z}^{-1}(q)$
for  $(a)_{\chic Z \chic Z}$),  a fact which allows their cancellation
to be actually implemented.
 
We now turn to the details of the calculation. 
Having right-handedly polarized fermions
is equivalent to saying that there are no contributions
coming from the graphs $(n)$, $(q)$, and $(u)+(v)$,
since these graphs vanish identically.
Next we show in detail how the cancellations 
of all GFP-dependent contributions take place.
The general strategy is to  
let the propagators on the left of each parentheses
hit there inverses inside the parentheses, and then
employ the identity of Eq.(\ref{xyz}). 
For example, the contributions from 
$(o)_{\gamma {\chic Z}}$ and $(p)_{{\chic Z} {\chic Z}}$ can be 
rewritten in the form
\bea
(o)_{\gamma {\chic Z}} &=& -{\cal V}_{\gamma {\chic f} \bar{\chic f}}^
{\mu}D_{\gamma} (q) \Bigg[s_w c_w \Bigg(
D_{\gamma}^{-1}(q) \bigg( \lambda_{\chic W}^2 I_4 k_{\mu}k_{\nu}
-2 \lambda_{\chic W} I_3  g_{\mu\nu} \bigg)
-2 \lambda_{\chic W} I_3 \bigg [k_{\mu}k_{\nu} -(k+q)^2g_{\mu\nu} 
\bigg ] \Bigg) D_{\chic Z}^{-1}(q)\Bigg]
\nonumber\\
&& D_{\chic Z}(q){\cal V}_{{\chic Z} \nu \bar{\nu}}^{\nu} \nonumber\\
(p)_{{\chic Z} {\chic Z}} &=& {\cal V}_{{\chic Z} 
{\chic f \bar{\chic f}}}^{\mu} D_{\chic Z} (q) 
\Bigg[c_w^2 \Bigg(D_{\chic Z}^{-1}(q) 
\bigg(\lambda_{\chic W}^2 I_4 k_{\mu}k_{\nu} - 2 \lambda_{\chic W} 
I_3 g_{\mu\nu} \bigg) + \lambda_{\chic W}^2 M_{\chic Z}^2  
I_4 k_{\mu}k_{\nu} - 2 \lambda_{\chic W} I_3 \bigg [ 
M_{\chic Z}^2 g_{\mu\nu} 
\nonumber\\ 
&& -(k+q)^2g_{\mu\nu} + k_{\mu}k_{\nu}\bigg ] \Bigg)
D_{\chic Z}^{-1}(q)\Bigg] D_{\chic Z}(q)
{\cal V}_{{\chic Z} \nu \bar{\nu}}^{\nu} 
\label{ogz}
\eea
and so in the case where the initial fermions are right-handed we have
\bea
(o)_{\gamma {\chic Z}}^{{\chic R} {\chic R}} 
+ (p)_{{\chic Z} {\chic Z}}^{{\chic R} {\chic R}} &=&
{\cal V}_{\gamma {\chic f_R} \bar{\chic f}_{\chic R}}^{\mu}D_{\gamma} (q)
\Bigg[s_w c_w \Bigg(2 \lambda_{\chic W} I_3 \bigg [ k_{\mu}k_{\nu}  
-(k+q)^2g_{\mu\nu} \bigg ] \Bigg)  D_{\chic Z}^{-1}(q)\Bigg]
D_{\chic Z}(q){\cal V}_{{\chic Z} \nu \bar{\nu}}^{\nu}\nonumber\\
&&
+{\cal V}_{\gamma {\chic f_R} \bar{\chic f}_{\chic R}}^{\mu}D_{\chic Z} (q)
\Bigg[s_w c_w \Bigg(\lambda_{\chic W}^2 M_{\chic Z}^2  I_4 k_{\mu}k_{\nu} 
 -2 \lambda_{\chic W} I_3 \bigg [ 
M_{\chic Z}^2 g_{\mu\nu} - (k+q)^2g_{\mu\nu} + k_{\mu}k_{\nu}\bigg ]
\Bigg) D_{\chic Z}^{-1}(q)\Bigg] \nonumber\\
&& D_{\chic Z}(q){\cal V}_{{\chic Z} \nu \bar{\nu}}^{\nu}
\label{opRR}
\eea
Similarly we obtain from the Abelian-like graphs of Fig.~7
\bea
[(s)+(t)]_{{\chic Z} {\chic Z}}^{{\chic R} {\chic R}} 
+ (r)_{\gamma {\chic Z}}^{{\chic R} {\chic R}} 
&=& - {\cal V}_{ \gamma {\chic f_R} \bar{\chic f}_{\chic R}}^{\mu} 
D_{\chic Z} (q) \Bigg[s_w c_w \lambda_{\chic W} D_{\chic Z}^{-1} (q)  
I_1 g_{\mu\nu}\Bigg] 
D_{\chic Z} (q) {\cal V}_{{\chic Z} \nu \bar{\nu}}^{\nu} \nonumber\\
&& + {\cal V}_{ \gamma {\chic f_R} \bar{\chic f}_{\chic R}}^{\mu} 
D_{\gamma} (q) \Bigg[s_w c_w \lambda_{\chic W} D_{\chic Z}^{-1} (q) 
I_1 g_{\mu\nu}\Bigg] D_{\chic Z} (q) {\cal V}_{{\chic Z} \nu \bar{\nu}}^{\nu} 
\label{stRR}
\eea
Here there is no cancellation except for the simplification
of writing the vertices in front using the relation
of Eq.(\ref{xyz}) .
Turning to the self-energy graphs we have
\bea
(a)_{{\chic Z} {\chic Z}}^{{\chic R} {\chic R}}  &=& 
(a)_{{\chic Z} {\chic Z}}^{{\chic R} {\chic R}\, (\xi_{\chic W}=1)} 
+ L_{{\chic Z} {\chic Z}}^{{\chic R} {\chic R}}(\xi_{\chic W}) 
+ R_{{\chic Z} {\chic Z}}^{{\chic R} {\chic R}}(\xi_{\chic W}) \nonumber\\ 
&& -{\cal V}_{{\chic Z} {\chic f} \bar{\chic f}}^{\mu} D_{\chic Z}(q)
\Bigg[c_w^2 \Bigg ( D_{\chic Z}^{-1}(q) S_{\mu\nu}
 + \lambda_{\chic W}^2 M_{\chic Z}^2 I_4 D_{\chic Z}^{-1} (q)  
k_{\mu}k_{\nu} \nonumber\\
&& - 2 \lambda_{\chic W} I_3 D_{\chic Z}^{-1}(q) 
\bigg (k_{\mu}k_{\nu} + M_{\chic Z}^2 g_{\mu\nu} 
- (k+q)^2 g_{\mu\nu}\bigg) \Bigg)\Bigg]
D_{\chic Z} (q) {\cal V}_{{\chic Z} \nu \bar{\nu}}^{\nu}
\nonumber\\ 
(a)_{\gamma {\chic Z}}^{{\chic R} {\chic R}}  &=& 
(a)_{\gamma {\chic Z}}^{{\chic R} {\chic R}\, (\xi_{\chic W}=1)} 
+ L_{\gamma {\chic Z}}^{{\chic R} {\chic R}} (\xi_{\chic W}) 
+ R_{\gamma {\chic Z}}^{{\chic R} {\chic R}} (\xi_{\chic W})
\nonumber\\ 
&& + {\cal V}_{ \gamma {\chic f} \bar{\chic f}}^{\mu} D_{\gamma}(q)
\Bigg[s_w c_w \Bigg( D_{\gamma}^{-1}(q) S_{\mu\nu}
 - 2 \lambda_{\chic W} I_3 
D_{\chic Z}^{-1}(q) \bigg \{ k_{\mu}k_{\nu} - (k+q)^2 g_{\mu\nu}\bigg \} 
\bigg]\Bigg)\Bigg]
D_{\chic Z} (q) {\cal V}_{{\chic Z} \nu \bar{\nu}}^{\nu} \nonumber\\
&& {}
\label{ExaZZ}
\eea
with
\be
S_{\mu\nu} = \lambda_{\chic W}^2 I_4 \bigg ( D_{\chic Z}^{-1}(q) 
  + M_{\chic Z}^2  \bigg )k_{\mu}k_{\nu} - 2 \lambda_{\chic W} I_3 
\bigg (D_{\chic Z}^{-1}(q) g_{\mu\nu} + k_{\mu}k_{\nu} 
+ M_{\chic Z}^2 g_{\mu\nu} - (k+q)^2 g_{\mu\nu} \bigg)
\label{Smunu}
\ee
Adding by parts we see that the terms proportional to $S_{\mu\nu}$ 
cancel, and we are left with
\bea
(a)_{{\chic Z} {\chic Z}}^{{\chic R} {\chic R}} 
+ (a)_{\gamma {\chic Z}}^{{\chic R} {\chic R}} &=& 
(a)_{{\chic Z} {\chic Z}}^{{\chic R} {\chic R}\, (\xi_{\chic W}=1)}  
+ (a)_{\gamma {\chic Z}}^{{\chic R} {\chic R}\, (\xi_{\chic W}=1)} +  
L_{{\chic Z} {\chic Z}}^{{\chic R} {\chic R}}(\xi_{\chic W}) 
+ R_{{\chic Z} {\chic Z}}^{{\chic R} {\chic R}}(\xi_{\chic W})
+ L_{\gamma {\chic Z}}^{{\chic R} {\chic R}}(\xi_{\chic W}) 
+ R_{\gamma {\chic Z}}^{{\chic R} {\chic R}}(\xi_{\chic W})
\nonumber\\
&& -{\cal V}_{\gamma {\chic f_R} \bar{\chic f}_{\chic R}}^{\mu} 
D_{\chic Z}(q) \Bigg[s_w c_w \Bigg (\lambda_{\chic W}^2 M_{\chic Z}^2 
I_4 D_{\chic Z}^{-1}(q)  k_{\mu}k_{\nu} \nonumber\\
&& - 2 \lambda_{\chic W} I_3 D_{\chic Z}^{-1}(q) \bigg 
(k_{\mu}k_{\nu} + M_{\chic Z}^2 g_{\mu\nu} -  (k+q)^2 g_{\mu\nu} \bigg) 
\Bigg)\Bigg] D_{\chic Z} (q) 
{\cal V}_{{\chic Z} \nu \bar{\nu}}^{\nu}\nonumber\\
&& -
{\cal V}_{\gamma {\chic f_R} \bar{\chic f}_{\chic R}}^{\mu} D_{\gamma}(q)
\Bigg[s_w c_w \Bigg ( 2\lambda_{\chic W} I_3 
\bigg \{ k_{\mu}k_{\nu} - (k+q)^2 g_{\mu\nu}\bigg \}D_{\chic Z}^{-1}(q) 
\bigg]\Bigg)\Bigg] D_{\chic Z} (q) {\cal V}_{{\chic Z} \nu \bar{\nu}}^{\nu}
\nonumber\\
&=& (a)_{{\chic Z} {\chic Z}}^{{\chic R} {\chic R}\, (\xi_{\chic W}=1)}  
(a)_{\gamma {\chic Z}}^{{\chic R} {\chic R}\, (\xi_{\chic W}=1)} +  
L_{{\chic Z} {\chic Z}}^{{\chic R} {\chic R}}(\xi_{\chic W}) 
+ R_{{\chic Z} {\chic Z}}^{{\chic R} {\chic R}}(\xi_{\chic W})
+ L_{\gamma {\chic Z}}^{{\chic R} {\chic R}}(\xi_{\chic W}) 
+ R_{\gamma {\chic Z}}^{{\chic R} {\chic R}}(\xi_{\chic W})
\nonumber\\
&& - \bigg ( (o)_{\gamma {\chic Z}}^{{\chic R} {\chic R}} 
+ (p)_{{\chic Z} {\chic Z}}^{{\chic R} {\chic R}} \bigg )
\label{ARR}
\eea

At this point it is clear that one has effectively arrived at the 
same stage one was in the previous section right after  Eq.(\ref{RGZLGZ}), 
except for the fact that now, instead of the $\xi_{\chic W}$-dependent
contributions from 
$[(s)+(t)]_{{\chic Z} {\chic Z}}$,  $(r)_{\gamma {\chic Z}}$,  
$((u)+(v))_{{\chic Z} {\chic Z}}$, $((u)+(v))_{\gamma {\chic Z}}$ in 
Eq.(\ref{Estruv}), we only have the $\xi_{\chic W}$-dependent
contributions from 
$[(s)+(t)]_{{\chic Z} {\chic Z}}^{{\chic R} {\chic R}}$ and 
$(r)_{\gamma {\chic Z}}^{{\chic R} {\chic R}}$. Notice however that, again 
by virtue of  Eq.(\ref{xyz}), now there is an extra cancellation possible 
between the part in 
$L_{{\chic Z} {\chic Z}}^{{\chic R} {\chic R}}(\xi_{\chic W}) 
+ R_{{\chic Z} {\chic Z}}^{{\chic R} {\chic R}}(\xi_{\chic W})$
proportional to $D_{\chic Z}^{-1}(q) I_1$ and the part in  
$L_{{\chic Z} {\chic Z}}^{{\chic R} {\chic R}}(\xi_{\chic W}) 
+ R_{{\chic Z} {\chic Z}}^{{\chic R} {\chic R}}(\xi_{\chic W})
+ L_{\gamma {\chic Z}}^{{\chic R} {\chic R}}(\xi_{\chic W}) 
+ R_{\gamma {\chic Z}}^{{\chic R} {\chic R}}(\xi_{\chic W})$
proportional to $D_{\gamma}^{-1}(q)I_1$, a fact which exactly compensates
for the absence of ($\xi_{\chic W}$-dependent) contributions from 
$((u)+(v))_{{\chic Z} {\chic Z}}$, $((u)+(v))_{\gamma {\chic Z}}$. 
Having realized that, then the rest of the cancellations
proceed exactly as in the unpolarized case.

\setcounter{equation}{0}
\section{ The gauge cancellations in the Background Field Method}

The BFM \cite{BFM1,Abb,AGS,PAG}
is a special gauge-fixing procedure which
preserves  the   symmetry    of  the  action   under   ordinary  gauge
transformations    with respect to  the   background (classical) gauge
field,  while  the  (quantum)   gauge fields appearing   in  the loops
transform homogeneously under the gauge group, i.e. as ordinary matter
fields which happened to   be assigned to the adjoint  representation.
As happens within   every gauge-fixing  scheme the off-shell   Green's
functions derived in this formalism depend explicitly  on the GFP, but
due the residual background symmetry they  have the advantage of being
gauge-invariant, i.e.  they satisfy naive  QED-like Ward identities to
all orders in perturbation theory, and for every value of the GFP.
As has been shown in detail, when the 
GFP-dependent BFM Green's function
are computed in the Feynman gauge they {\it coincide}
with the GFP-independent PT  Green's function, both
at one loop \cite{BFMPT} and at two loops \cite{PT2L}.

In this section we will demonstrate  that the GFP cancellations in the
context of the  BFM proceed exactly  as  in the case  of the $R_{\xi}$
gauges, i.e.  through algebraic  manipulations only, provided that one
exploits again the fundamental cancellation mechanism operating in the
$S$-matrix  element; the latter  holds regardless  of the gauge-fixing
scheme chosen  when quantizing the theory.  It turns out that once the
GFP cancellations have been  implemented one recovers exactly the same
set of Feynman diagrams  one had ended up with  in the context of  the
$R_{\xi}$ gauges once the corresponding GFP cancellation procedure had
been completed, i.e.   at  the end  of section  3.  Evidently  the GFP
cancellation carried   out using  the  PT rules   projects us from one
gauge-fixing scheme   (BFM)  to   another  ($R_{\xi}$).    This  makes
abundantly  clear  the  fact  that  the  BFM  Feynman gauge  (which is
different from the $R_{\xi}$  Feynman gauge) has  not been singled out
in any a-priori manner, since, in fact, one is  effectively led to the
Feynman rules of a different gauge fixing scheme.  Of course, one will
reach eventually   the  BFM in  the Feynman   gauge (regardless of the
starting point) once the PT  procedure has been  completed, i.e.  once
the final rearrangement, described in the  following section, has been
carried out.

In the BFM the elementary three-boson vertex 
between a background field $\widetilde {A}_{\mu} (q)$
and two quantum fields $A_{\rho}(p_2)$ 
and $A_{\sigma} (p_3)$ for a general value of the 
quantum gauge-fixing parameter $\xi^{\chic W}_{\chic Q}$
reads \cite{DDW}
\be
{\widetilde {\Gamma}}_{\mu\rho\sigma}(q,p_2,p_3)
= \Bigg( q-p_2-\frac{1}{\xi^{\chic W}_{\chic Q}}\, p_3 
\Bigg)_{\sigma} g_{\mu\rho} + (p_2-p_3)_{\mu} g_{\rho\sigma} 
+ \Bigg(p_3-q + \frac{1}{\xi^{\chic W}_{\chic Q}}\, p_2 
\Bigg)_{\rho}g_{\mu\sigma} \, .
\label{GVerBfm}
\ee
which, inside the loop can be written as
\be
{\widetilde {\Gamma}}_{\mu\rho\sigma}(q,k,-k-q)
= \Gamma_{\mu\rho\sigma}^{\chic F} - 
\Bigg(\frac{1- {\xi^{\chic W}_{\chic Q}}} 
{{\xi^{\chic W}_{\chic Q}}} \Bigg)
\Gamma_{\mu\rho\sigma}^{\chic P} 
\label{TGbfm}
\ee
or
\be
{\widetilde {\Gamma}}_{\mu\rho\sigma}(q,k,-k-q)=
\Gamma_{\mu\rho\sigma}(q,k,-k-q) - 
\frac{1}{\xi^{\chic W}_{\chic Q}} \Gamma_{\mu\rho\sigma}^{\chic P}
\label{orTG}
\ee
In what follows we will employ the latter decomposition.
This will allow to project a large part of the calculation
into effectively the $R_{\xi}$ diagrammatic formulation and 
expedite the demonstration of the cancellations by
using directly the results of the previous section,
by simply setting  $\xi \to \xi^{\chic W}_{\chic Q}$.  
Similarly, the elementary four-gauge-boson vertex 
involving two background fields, $\widetilde {A}_{\mu}$ and
$\widetilde {A}_{\nu}$, and two quantum fields, $A_{\rho}$ 
and $A_{\sigma}$, reads
\bea
{\widetilde {\Gamma}}_{\mu\nu\rho\sigma} &=&
2 g_{\mu\nu}g_{\rho\sigma} + 
\bigg(\frac{1- \xi^{\chic W}_{\chic Q}}{\xi^{\chic W}_{\chic Q}}\bigg) 
g_{\mu\sigma}g_{\nu\rho}  
+ \bigg(\frac{1 - \xi^{\chic W}_{\chic Q}}{\xi^{\chic W}_{\chic Q}}\bigg) 
g_{\mu\rho}g_{\nu\sigma} \nonumber \\ 
&=& {\Gamma}_{\mu\nu\rho\sigma} + \frac{1}{\xi^{\chic W}_{\chic Q}}
\big( g_{\mu\sigma}g_{\nu\rho} + g_{\mu\rho}g_{\nu\sigma} \big)
\eea
Finally, the tree-level gauge boson propagators 
$\widetilde {\Delta}^{\mu\nu}_{\tilde{V}} (k)$ are identical to those 
given in Eq.(\ref{GenProp}), with $\xi \to \xi^{\chic W}_{\chic Q}$.
The strategy of the proof consists in splitting all BFM Feynman graphs as 
a part which is identical to the corresponding Feynman graph in the 
$R_{\xi}$ gauge, plus a leftover $\xi^{\chic W}_{\chic Q}$-dependent
contribution. Then, all one has to show is that all such additional 
contribution cancel against each other.

To prove that this is indeed the case, let us first list the graphs which 
are identical to those in the $R_{\xi}$ after the trivial change 
$\xi \to \xi^{\chic W}_{\chic Q}$:
\bea
(\widetilde{n}) &=& (n) \nonumber\\
(\widetilde{r}) &=& (r) \nonumber\\
(\widetilde{s}) + (\widetilde{t}) &=& (s)+(t) \nonumber\\
(\widetilde{u}) + (\widetilde{v}) &=& (u)+(v) \nonumber\\
(\widetilde{f}) &=& (f) \nonumber\\
(\widetilde{h}) &=& (h) \nonumber\\
(\widetilde{i}) &=& (i) \nonumber\\
(\widetilde{j}) &=& (j) \nonumber\\
(\widetilde{k}) &=& (k) \nonumber\\
\label{tilnr}
\eea

The rest of the diagrams differ from those in the $R_{\xi}$ gauge.
In particular, for the vertex graphs we have
\bea
(\widetilde{o}) &=&  (o) + (\widetilde{o})_{\gamma {\chic Z}}  \nonumber\\
(\widetilde{p}) &=&  (p) + (\widetilde{p})_{{\chic Z} {\chic Z}}  \nonumber\\
(\widetilde{q}) &=&  (q) + (\widetilde{q})_{\gamma {\chic Z}} 
+ (\widetilde{q})_{{\chic Z} {\chic Z}}  
\label{tilqp}
\eea
with
\bea
(\widetilde{o})_{\gamma {\chic Z}} &=& 
-{\cal V}_{\gamma {\chic f} \bar{\chic f}}^{\mu}D_{\gamma} (q)
\Bigg[2s_w c_w I_5 \Delta_{\mu\nu}^{\chic W}(k+q) D_{\chic Z}^{-1}(q)\Bigg]
D_{\chic Z}(q){\cal V}_{{\chic Z} \nu\bar{\nu}}^{\nu} \nonumber\\
(\widetilde{p})_{{\chic Z} {\chic Z}} &=& 
{\cal V}_{{\chic Z} {\chic f} \bar{\chic f}}^{\mu}D_{\chic Z} (q)
\Bigg[2c_w^2 I_5 \Delta_{\mu\nu}^{\chic W}(k+q)D_{\chic Z}^{-1}(q)\Bigg] 
D_{\chic Z}(q){\cal V}_{{\chic Z} \nu\bar{\nu}}^{\nu} \nonumber\\
(\widetilde{q})_{\gamma {\chic Z}} &=& 
-{\cal V}_{\gamma {\chic f} \bar{\chic f}}^{\mu}D_{\gamma} (q)
\Bigg[2s_w c_w D_{\gamma}^{-1}(q) I_5 \Delta_{\mu\nu}^{\chic W}(k+q)\Bigg]
D_{\chic Z} (q) {\cal V}_{{\chic Z} \nu \bar{\nu}}^{\nu}
\nonumber\\
(\widetilde{q})_{{\chic Z} {\chic Z}} &=& 
{\cal V}_{{\chic Z} {\chic f} \bar{\chic f}}^{\mu} D_{\chic Z}(q)
\Bigg[2 c_w^2 D_{\chic Z}^{-1}(q) I_5 \Delta_{\mu\nu}^{\chic W}(k+q)\Bigg]
D_{\chic Z} (q) {\cal V}_{{\chic Z} \nu \bar{\nu}}^{\nu}
\label{TilGZ}
\eea
Similarly, the self-energy graphs give
\bea
(\widetilde{a})_{{\chic Z} {\chic Z}} &=& 
(a)_{{\chic Z} {\chic Z}} + G_{{\chic Z} {\chic Z}} 
(\xi^{\chic W}_{\chic Q})\nonumber\\
(\widetilde{a})_{\gamma {\chic Z}} &=& 
(a)_{\gamma {\chic Z}} + G_{\gamma {\chic Z}}
(\xi^{\chic W}_{\chic Q})
\label{atil}
\eea
with
\bea 
G_{{\chic Z} {\chic Z}} (\xi^{\chic W}_{\chic Q}) &=& 
{\cal V}_{{\chic Z} {\chic f} \bar{\chic f}}^{\mu}D_{\chic Z} (q)
\Bigg[2c_w^2 \Bigg( 2 D_{\chic Z}^{-1}(q) I_5 \Delta_{\mu\nu}^{\chic W}(k+q)
+ 3 I_2 k_{\mu}k_{\nu}  + 
\bigg(\frac{1}{\xi^{\chic W}_{\chic Q}}\bigg)\Delta_{\mu\nu}^{\chic W}(k) 
\nonumber\\
&& -2 I_5 g_{\mu\nu} + (2 M_{\chic Z}^{2} - M_{\chic W}^{2}) 
I_5 \Delta_{\mu\nu}^{\chic W}(k+q) \Bigg) \Bigg] 
D_{\chic Z}(q){\cal V}_{{\chic Z} \nu\bar{\nu}}^{\nu} \nonumber\\
G_{\gamma {\chic Z}}(\xi^{\chic W}_{\chic Q}) &=&
-{\cal V}_{ \gamma {\chic f} \bar{\chic f}}^{\mu}D_{\gamma} (q)
\Bigg[2 s_w c_w \Bigg( \bigg\{D_{\gamma}^{-1}(q) + D_{\chic Z}^{-1}(q) 
\bigg\} I_5 \Delta_{\mu\nu}^{\chic W}(k+q) + 3 I_2 k_{\mu}k_{\nu}  + 
\bigg(\frac{1}{\xi^{\chic W}_{\chic Q}}\bigg)\Delta_{\mu\nu}^{\chic W}(k) 
\nonumber\\
&& -2 I_5 g_{\mu\nu} + (M_{\chic Z}^{2} - M_{\chic W}^{2}) 
I_5 \Delta_{\mu\nu}^{\chic W}(k+q) \Bigg) \Bigg] 
D_{\chic Z}(q){\cal V}_{{\chic Z} \nu\bar{\nu}}^{\nu} \nonumber\\
\label{GGZ}
\eea
and the remaining self-energy graphs yield
\bea
(\widetilde{b})_{{\chic Z} {\chic Z}} 
+ (\widetilde{c})_{{\chic Z} {\chic Z}}  &=& (b)_{{\chic Z} {\chic Z}} 
+ (c)_{{\chic Z} {\chic Z}} + {\cal V}_{{\chic Z} {\chic f} 
\bar{\chic f}}^{\mu}D_{\chic Z} (q)
\Bigg[2c_w^2 (2 M_{\chic Z}^{2} - M_{\chic W}^{2}) I_5 
\Delta_{\mu\nu}^{\chic W}(k+q) \Bigg] D_{\chic Z}(q)
{\cal V}_{{\chic Z} \nu\bar{\nu}}^{\nu}
\nonumber\\
(\widetilde{d})_{{\chic Z} {\chic Z}} 
+ (\tilde{e})_{{\chic Z} {\chic Z}}  &=& (d)_{{\chic Z} {\chic Z}} 
+ (e)_{{\chic Z} {\chic Z}} 
+ {\cal V}_{{\chic Z} {\chic f} \bar{\chic f}}^{\mu}D_{\chic Z} (q)
\Bigg[6 c_w^2 I_2 k_{\mu}k_{\nu}\Bigg] 
D_{\chic Z}(q){\cal V}_{{\chic Z} \nu \bar{\nu}}^{\nu}
\nonumber\\
(\widetilde{g})_{{\chic Z} {\chic Z}} &=&  (g)_{{\chic Z} {\chic Z}} +  
{\cal V}_{{\chic Z} {\chic f} \bar{\chic f}}^{\mu}D_{\chic Z} (q)
\Bigg[2 c_w^2 \bigg(\frac{1}{\xi^{\chic W}_{\chic Q}}\bigg)
\Delta_{\mu\nu}^{\chic W}(k) \Bigg] D_{\chic Z}(q) 
{\cal V}_{{\chic Z} \nu\bar{\nu}}^{\nu}
\nonumber\\
(\widetilde{\ell})_{{\chic Z} {\chic Z}}  &=& 
{\cal V}_{{\chic Z} {\chic f} \bar{f}}^{\mu}D_{\chic Z} (q)\Bigg[4 c_w^2
I_5 g_{\mu\nu} \Bigg] D_{\chic Z}(q){\cal V}_{{\chic Z} \nu\bar{\nu}}^{\nu}
\label{lgtil}
\eea
and
\bea
(\widetilde{b})_{\gamma {\chic Z}} + (\widetilde{c})_{\gamma {\chic Z}} 
&=& 0 = (b)_{\gamma {\chic Z}} + (c)_{\gamma {\chic Z}}  + 
{\cal V}_{\gamma {\chic f} \bar{\chic f}}^{\mu}D_{\gamma} (q)
\Bigg[2 s_w c_w (M_{\chic Z}^{2} - M_{\chic W}^{2}) I_5 
\Delta_{\mu\nu}^{\chic W}(k+q) \Bigg] D_{\chic Z}(q) 
{\cal V}_{{\chic Z} \nu \bar{\nu}}^{\nu}
\nonumber\\
(\widetilde{d})_{\gamma {\chic Z}} + (\widetilde{e})_{\gamma {\chic Z}} 
&=& (d)_{\gamma {\chic Z}} + (e)_{\gamma {\chic Z}}  
+{\cal V}_{\gamma {\chic f} \bar{\chic f}}^{\mu} D_{\gamma} (q)
\Bigg[6 s_w c_w I_2 k_{\mu}k_{\nu}\Bigg] 
D_{\chic Z}(q){\cal V}_{{\chic Z} \nu\bar{\nu}}^{\nu}
\nonumber\\
(\widetilde{g})_{\gamma {\chic Z}} &=&  (g)_{\gamma {\chic Z}} +
{\cal V}_{\gamma {\chic f} \bar{\chic f}}^{\mu}D_{\gamma} (q)
\Bigg[2 s_w c_w \bigg(\frac{1}{\xi^{\chic W}_{\chic Q}} \bigg)
\Delta_{\mu\nu}^{\chic W}(k) \Bigg] D_{\chic Z}(q)
{\cal V}_{{\chic Z} \nu\bar{\nu}}^{\nu}
\label{dbtil}
\eea
Adding     Eq.(\ref{tilnr})-Eq.(\ref{dbtil}),     by   parts    it  is
straightforward to verify that all additional contributions cancel and
one recovers simply the entire  $S$-matrix element, expressed now 
as a sum
of   the  $R_\xi$-gauge  Feynman diagrams.   Notice    again that no
integration over virtual momenta have been carried out .

To see how the above cancellation proceeds in the polarized case,
simply notice that in that case (i) 
$(\widetilde{q})_{\gamma {\chic Z}} = 
(\widetilde{q})_{{\chic Z} {\chic Z}} = 0$ since
the diagram $(q)$ is identically zero in this case; (ii)
half of the term in $G_{{\chic Z} {\chic Z}} (\xi^{\chic W}_{\chic Q})$ 
proportional to $D_{\chic Z}^{-1}(q)$ and the term in 
$G_{\gamma {\chic Z}} (\xi^{\chic W}_{\chic Q})$ proportional to 
$D_{\gamma}^{-1}(q)$ cancel against each other; (iii)
finally, the remaining term in $G_{{\chic Z} {\chic Z}} 
(\xi^{\chic W}_{\chic Q})$ 
proportional to $D_{\chic Z}^{-1}(q)$ cancels against 
$(\widetilde{p})_{{\chic Z} {\chic Z}}$ and the term 
in $G_{\gamma {\chic Z}} (\xi^{\chic W}_{\chic Q})$ proportional to 
$D_{\chic Z}^{-1}(q)$ cancels against $(\widetilde{o})_{\gamma {\chic Z}}$.
The rest of the cancellation proceeds as in the unpolarized case.

The construction presented  above is  in a  way complementary to  that
presented in \cite{GPT}; there it was shown that  the judicious use of
tree-level WI allows one to to start from the Feynman diagrams written
in  the  $R_{\xi}$ gauges  and project  oneself   into the  BFM set of
Feynman  diagrams   for {\it  any} choice  of  the  BFM GFP.  Here the
reverse has  been shown: one  may start from the  Feynman rules of the
BFM (in any GFP)  and, by eliminating longitudinal  momenta, end up in
the   $R_{\xi}$  gauges.   Notice     that  the   freedom  in   moving
diagrammatically from one  gauge fixing scheme  to  the next does  not
mean that all gauges are   physically equivalent; indeed, as has  been
explained in detail  in the literature \cite{PP2,PRW},  it  is only in
the context of the PT, i.e.  once {\it all} (and not only some) of the
longitudinal momenta have been eliminated by virtue of the WI that one
arrives at a     physically meaningful  separation  of   the  original
amplitude into self-energy, vertex, and box-like sub-amplitudes. 

\setcounter{equation}{0}
\section{ The final rearrangement}

It is known that, when the remaining longitudinal momenta
stemming from the decomposition of the vertex described in 
Eq.(\ref{decomp}) - Eq.(\ref{GFGP})
are allowed to trigger the same WI as their counterparts
originating from the tree-level propagators, the final one-loop 
answer reorganizes itself further into effective one-loop
self-energies,vertices
and boxes which, in addition to being GFP-independent, are
gauge-invariant, i.e. they satisfy naive, QED-like WI 
\cite{PTCPT,PT,PT2}.
To see that one 
simply employs the vertex decomposition given in 
Eq.(\ref{decomp}) - Eq.(\ref{GFGP}) inside vertex diagrams
containing a tri-linear vertex,
and reassigns the propagator-like contributions which emerge
to the conventional self-energy graphs, thus constructing the
one-loop PT self-energy $\widehat{\Pi}_{\mu\nu}$.  
The remaining genuine vertex graphs
constitute the effective one-loop PT vertex 
$\widehat{\Gamma}_{\mu}$. 

In what follows we will use hats to denote the individual PT Feynman 
diagrams emerging after this final rearrangement has been carried
out. Clearly,
\bea
(\widehat{n}) &=& (n)_{(\xi_{\chic W} =1)} 
= (\widetilde{n})_{(\xi^{\chic W}_{\chic Q} =1)}
\nonumber\\
(\widehat{r}) &=& (r)_{(\xi_{\chic W} =1)} 
= (\widetilde{r})_{(\xi^{\chic W}_{\chic Q} =1)}
\nonumber\\
(\widehat{s}) &=& (s)_{(\xi_{\chic W} =1)} 
= (\widetilde{s})_{(\xi^{\chic W}_{\chic Q} =1)}
\nonumber\\
(\widehat{t}) &=& (t)_{(\xi_{\chic W} =1)} 
= (\widetilde{t})_{(\xi^{\chic W}_{\chic Q} =1)}
\nonumber\\
(\widehat{u}) &=& (u)_{(\xi_{\chic W} =1)} 
= (\widetilde{u})_{(\xi^{\chic W}_{\chic Q} =1)}
\nonumber\\
(\widehat{v}) &=& (v)_{(\xi_{\chic W} =1)} 
= (\widetilde{v})_{(\xi^{\chic W}_{\chic Q} =1)}
\label{hatnr}
\eea
The first relation in the above equation simply states that
the PT box is equal to the conventional one in the 
Feynman gauge, both in the $R_{\xi}$ gauges and in the BFM.

From the vertex graphs containing tri-linear vertices we have:
\bea
(\widehat{o}) &=& (o)_{(\xi_{\chic W} =1)} -  
 (o)_{\gamma {\chic Z}}^{\Gamma^P}
= (\widetilde{o})_{(\xi^{\chic W}_{\chic Q} =1)} 
\nonumber\\
(\widehat{p}) &=& (p)_{(\xi_{\chic W} =1)} -  
 (p)_{{\chic Z} {\chic Z}}^{\Gamma^P} 
= (\widetilde{p})_{(\xi^{\chic W}_{\chic Q} =1)} 
\nonumber\\
(\widehat{q}) &=& (q)_{(\xi_{\chic W} =1)} -  
 \bigg \{ (q)_{{\chic Z} {\chic Z}}^{\Gamma^P} 
+ (q)_{\gamma {\chic Z}}^{\Gamma^P}\bigg \}
= (\widetilde{q})_{(\xi^{\chic W}_{\chic Q} =1)}
\label{hatopq}
\eea
where
\bea
(o)_{\gamma {\chic Z}}^{\Gamma^P}&=&
-2 {\cal V}_{\gamma {\chic f} \bar{\chic f}}^{\mu}D_{\gamma} (q)
\Big[s_w c_w I_0 D_{\chic Z}^{-1}(q)\Big]
D_{\chic Z}(q){\cal V}_{{\chic Z} \nu \bar{\nu}}^{\nu} \nonumber\\
(p)_{{\chic Z} {\chic Z}}^{\Gamma^P} &=&
2 {\cal V}_{{\chic Z} {\chic f} \bar{\chic f}}^{\mu}D_{\chic Z} (q)
\Big[c_w^2 I_0 D_{\chic Z}^{-1}(q)\Big] 
D_{\chic Z}(q){\cal V}_{{\chic Z} \nu \bar{\nu}}^{\nu} \nonumber\\
(q)_{{\gamma} {\chic Z}}^{\Gamma^P} &=&
-2 {\cal V}_{\gamma {\chic f} \bar{\chic f}}^{\mu}D_{\gamma} (q)
\Big[s_w c_w D_{\gamma}^{-1}(q) I_0 \Big]
D_{\chic Z} (q) {\cal V}_{{\chic Z} \nu \bar{\nu}}^{\nu}
\nonumber\\
(q)_{ {\chic Z} {\chic Z}}^{\Gamma^P} &=&
2 {\cal V}_{{\chic Z} {\chic f} \bar{\chic f}}^{\mu} D_{\chic Z}(q)
\Big[c_w^2 D_{\chic Z}^{-1}(q)  I_0 \Big]
D_{\chic Z} (q) {\cal V}_{{\chic Z} \nu \bar{\nu}}^{\nu}
\label{GPGZ}
\eea
are the propagator-like parts emerging after the longitudinal
momenta composing $\Gamma^P_{\alpha\mu\nu}$ (see Eq.(\ref{GFGP}))
have triggered the WI of  Eq.(\ref{EWI}). 
Evidently the effective proper one-loop photon-neutrino
vertex $\widehat{\Gamma}^{\mu}_{\gamma \nu \bar{\nu}}$
and the PT one-loop $Z$-neutrino vertex
$\widehat{\Gamma}^{\mu}_{{\chic Z} \nu \bar{\nu}}$
are given by 
\bea
{\cal V}_{\gamma {\chic f} \bar{\chic f}}^{\nu} g_{\nu\mu} D_{\gamma}(q)
\widehat{\Gamma}^{\mu}_{\gamma \nu \bar{\nu}} 
&=& (\widetilde{o})_{(\xi^{\chic W}_{\chic Q} =1)} 
+ (\widetilde{r})_{(\xi^{\chic W}_{\chic Q} =1)}
\nonumber\\
{\cal V}_{{\chic Z} {\chic f} \bar{\chic f}}^{\nu} g_{\nu\mu} D_{\chic Z}(q)
\widehat{\Gamma}^{\mu}_{{\chic Z} \nu \bar{\nu}} 
&=& (\widetilde{p})_{(\xi^{\chic W}_{\chic Q} =1)} + 
(\widetilde{s})_{(\xi^{\chic W}_{\chic Q} =1)} 
+ (\widetilde{t})_{(\xi^{\chic W}_{\chic Q} =1)} 
\label{nuevo}
\eea

Finally, the PT self-energies
$\widehat{\Pi}_{\mu\nu}^{{\chic Z} {\chic Z}}(q)$ and 
$\widehat{\Pi}_{\mu\nu}^{\gamma {\chic Z}}(q)$ are simply 
the sum of all remaining self-energy-like 
contributions, i.e. 
\bea
{\cal V}_{{\chic Z} {\chic f} \bar{\chic f}}^{\mu} D_{\chic Z}(q)\Bigg[
\widehat{\Pi}_{\mu\nu}^{{\chic Z}{\chic Z}}(q)\Bigg] D_{\chic Z}(q)
{\cal V}_{{\chic Z} \nu \bar{\nu}}^{\nu} &=& 
{\cal V}_{{\chic Z} {\chic f} \bar{\chic f}}^{\mu} D_{\chic Z}(q)\Bigg[
\Pi_{\mu\nu}^{{\chic Z}{\chic Z}}(q,\xi_{\chic W}=1)\Bigg] D_{\chic Z}(q)
{\cal V}_{{\chic Z} \nu \bar{\nu}}^{\nu} +  
(p)_{{\chic Z}{\chic Z}}^{\Gamma^P} + 
(q)_{{\chic Z}{\chic Z}}^{\Gamma^P} 
\nonumber\\
{\cal V}_{\gamma {\chic f} \bar{\chic f}}^{\mu} D_{\gamma}(q)\Bigg[
\widehat{\Pi}_{\mu\nu}^{{\chic Z}{\chic Z}}(q)\Bigg]D_{\chic Z}(q)
{\cal V}_{{\chic Z} \nu \bar{\nu}}^{\nu} &=& 
{\cal V}_{\gamma {\chic f} \bar{\chic f}}^{\mu} D_{\gamma}(q)\Bigg[
\Pi_{\mu\nu}^{\gamma {\chic Z}}(q,\xi_{\chic W}=1)\Bigg]D_{\chic Z}(q)
{\cal V}_{{\chic Z} \nu \bar{\nu}}^{\nu} +  
(o)_{\gamma {\chic Z}}^{\Gamma^P} + 
(q)_{\gamma {\chic Z}}^{\Gamma^P} \nonumber\\
&&{}
\label{VmuZff}
\eea
It is a matter of straightforward algebra to prove that
in fact
\bea
\widehat{\Pi}_{\mu\nu}^{{\chic Z} {\chic Z}}(q) &=& 
\widetilde{\Pi}_{\mu\nu}^{{\chic Z} {\chic Z}}
(q, \xi^{\chic W}_{\chic Q} =1) \nonumber\\
\widehat{\Pi}_{\mu\nu}^{\gamma {\chic Z}}(q) &=& 
\widetilde{\Pi}_{\mu\nu}^{ \gamma {\chic Z}}
(q, \xi^{\chic W}_{\chic Q} =1)
\label{hatPi}
\eea
To see that in detail, use the identity
\be
 \Gamma_{\mu\rho\sigma}\Gamma_{\nu}^{\rho\sigma}
=
\Gamma_{\mu\rho\sigma}^{\chic F}\Gamma_{\nu}^{{\chic F} \rho\sigma}
+ \Gamma_{\mu\rho\sigma}^{\chic P}\Gamma_{\nu}^{\rho\sigma} 
+ \Gamma_{\mu\rho\sigma}\Gamma_{\nu}^{{\chic P} \rho\sigma} 
- \Gamma_{\mu\rho\sigma}^{\chic P}\Gamma_{\nu}^{{\chic P} \rho\sigma}
\label{Gmurho}
\ee
to cast the diagrams $({a})_{{\chic Z} {\chic Z}}^{(\xi_{\chic W} =1)}$ 
and $(a)_{\gamma {\chic Z}}^{(\xi_{\chic W} =1)}$ in the form
\bea
({a})_{{\chic Z} {\chic Z}}^{(\xi_{\chic W} =1)} &=& 
(\widetilde{a})_{{\chic Z} {\chic Z}}^{(\xi^{\chic W}_{\chic Q} =1)}
+ G_{{\chic Z} {\chic Z}}^{\Gamma^P} \nonumber\\
({a})_{\gamma {\chic Z}}^{(\xi_{\chic W} =1)} &=& 
(\widetilde{a})_{\gamma {\chic Z}}^{(\xi^{\chic W}_{\chic Q} =1)}
+ G_{\gamma {\chic Z}}^{\Gamma^P}
\label{xi1aZZ}
\eea
with
\bea 
G_{{\chic Z} {\chic Z}}^{\Gamma^P} &=& 
- {\cal V}_{{\chic Z} {\chic f} \bar{\chic f}}^{\mu}D_{\chic Z} (q)
\Bigg[2c_w^2 \Bigg( 2 D_{\chic Z}^{-1}(q) I_0 g_{\mu\nu}
+ 3 I_0 k_{\mu}k_{\nu}   
- D_{\chic W} (k) g_{\mu\nu} + (2 M_{\chic Z}^{2} - M_{\chic W}^{2}) I_0 
\Bigg) \Bigg] D_{\chic Z}(q)
{\cal V}_{{\chic Z} \nu\bar{\nu}}^{\nu} \nonumber\\
G_{\gamma {\chic Z}}^{\Gamma^P}  &=&
{\cal V}_{ \gamma {\chic f} \bar{\chic f}}^{\mu}D_{\gamma} (q)
\Bigg[2 s_w c_w \Bigg( \bigg\{
D_{\gamma}^{-1}(q) + D_{\chic Z}^{-1}(q) \bigg\} I_0 g_{\mu\nu}
+ 3 I_0 k_{\mu}k_{\nu}   
 - D_{\chic W} (k) g_{\mu\nu} + (M_{\chic Z}^{2} 
- M_{\chic W}^{2}) I_0 g_{\mu\nu} \Bigg) \Bigg] \nonumber\\
&& D_{\chic Z}(q){\cal V}_{{\chic Z} \nu\bar{\nu}}^{\nu} 
\label{GGPZZ}
\eea
Note that $G_{{\chic Z} {\chic Z}}^{\Gamma^P} = 
G_{{\chic Z} {\chic Z}} (\xi^{\chic W}_{\chic Q} =1)$ 
and $G_{\gamma {\chic Z}}^{\Gamma^P} 
=  G_{\gamma {\chic Z}}(\xi^{\chic W}_{\chic Q} =1)$ 
Then, simple algebra yields for the remaining graphs:
\bea
(b)_{{\chic Z} {\chic Z}}^{(\xi_{\chic W} =1)} 
+ (c)_{{\chic Z} {\chic Z}}^{(\xi_{\chic W} =1)}
&=& 
 (\widetilde{b})_{{\chic Z} {\chic Z}}^{(\xi^{\chic W}_{\chic Q} =1)} + 
(\widetilde{c})_{{\chic Z} {\chic Z}}^{(\xi^{\chic W}_{\chic Q} =1)} 
+ {\cal V}_{{\chic Z} {\chic f} \bar{\chic f}}^{\mu} D_{\chic Z} (q)
\Bigg[2c_w^2 (2 M_{\chic Z}^{2} - M_{\chic W}^{2}) I_0 g_{\mu\nu}
\Bigg] D_{\chic Z}(q){\cal V}_{{\chic Z} \nu\bar{\nu}}^{\nu}
\nonumber\\
(d)_{{\chic Z} {\chic Z}}^{(\xi_{\chic W} =1)}
+ (e)_{{\chic Z} {\chic Z}}^{(\xi_{\chic W} =1)}
&=& (\widetilde{d})_{{\chic Z} {\chic Z}}^{(\xi^{\chic W}_{\chic Q} =1)} 
+ (\widetilde{e})_{{\chic Z} {\chic Z}}^{(\xi^{\chic W}_{\chic Q} =1)} 
+ {\cal V}_{{\chic Z} {\chic f} \bar{\chic f}}^{\mu}D_{\chic Z} (q)
\Bigg[6 c_w^2 I_0 k_{\mu}k_{\nu}\Bigg] 
D_{\chic Z}(q){\cal V}_{{\chic Z} \nu \bar{\nu}}^{\nu}
\nonumber\\
(g)_{{\chic Z} {\chic Z}}^{(\xi_{\chic W} =1)}  &=& 
(\widetilde{g})_{{\chic Z} {\chic Z}}^{(\xi^{\chic W}_{\chic Q} =1)}
-{\cal V}_{{\chic Z} {\chic f} \bar{\chic f}}^{\mu} D_{\chic Z} (q)
\Bigg[2 c_w^2 D_{\chic W}(k) \Bigg] 
D_{\chic Z}(q){\cal V}_{{\chic Z} \nu \bar{\nu}}^{\nu}
\nonumber\\
\label{detil}
\eea
and
\bea
(b)_{\gamma {\chic Z}}^{(\xi_{\chic W} =1)} 
+ (c)_{\gamma {\chic Z}}^{(\xi_{\chic W} =1)}
&=& = - {\cal V}_{\gamma {\chic f} \bar{\chic f}}^{\mu}D_{\gamma} (q)
\Bigg[2 s_w c_w (M_{\chic Z}^{2} - M_{\chic W}^{2}) I_0 g_{\mu\nu}
\Bigg] D_{\chic Z}(q){\cal V}_{{\chic Z} \nu \bar{\nu}}^{\nu}
\nonumber\\
(d)_{\gamma {\chic Z}}^{(\xi_{\chic W} =1)} 
+ (e)_{\gamma {\chic Z}}^{(\xi_{\chic W} =1)} 
&=& (\widetilde{d})_{\gamma {\chic Z}}^{(\xi^{\chic W}_{\chic Q} =1)}  + 
(\widetilde{e})_{\gamma {\chic Z}}^{(\xi^{\chic W}_{\chic Q} =1)}
+ {\cal V}_{\gamma {\chic f} \bar{\chic f}}^{\mu} D_{\gamma} (q)
\Bigg[6 s_w c_w I_0 k_{\mu}k_{\nu}\Bigg] 
D_{\chic Z}(q){\cal V}_{{\chic Z} \nu\bar{\nu}}^{\nu}
\nonumber\\
(g)_{\gamma {\chic Z}}^{(\xi_{\chic W} =1)} &=& 
(\widetilde{g})_{\gamma {\chic Z}}^{(\xi^{\chic W}_{\chic Q} =1)} +
{\cal V}_{\gamma {\chic f} \bar{\chic f}}^{\mu}D_{\gamma} (q)
\Bigg[2 s_w c_w 
D_{\chic W}(k) \Bigg] D_{\chic Z}(q){\cal V}_{{\chic Z} \nu\bar{\nu}}^{\nu}
\label{gdtil}
\eea
Summing by parts Eq.(\ref{xi1aZZ}) and Eq.(\ref{detil}) - Eq.(\ref{gdtil}), 
separating the $ZZ$ from the $\gamma Z$ components 
one arrives at Eq.(\ref{hatPi}).

The final rearrangement in the polarized case proceeds 
as follows:
(i) $(q)_{ {\chic Z} {\chic Z}}^{\Gamma^P}=
(q)_{ {\gamma} {\chic Z}}^{\Gamma^P}=0$; 
(ii) half of the term in $G_{{\chic Z} {\chic Z}}^{\Gamma^P}$
proportional to $D_{\chic Z}^{-1}(q)$ and the term in 
$G_{\gamma {\chic Z}}^{\Gamma^P}$
proportional to 
$D_{\gamma}^{-1}(q)$ cancel against each other; (iii)
the remaining term in $G_{{\chic Z} {\chic Z}}^{\Gamma^P}$
proportional to $D_{\chic Z}^{-1}(q)$ cancels against 
$(p)_{{\chic Z}{\chic Z}}^{\Gamma^P}$, 
and the term 
in $G_{\gamma {\chic Z}}^{\Gamma^P}$
proportional to 
$D_{\chic Z}^{-1}(q)$ cancels against 
$(o)_{\gamma {\chic Z}}^{\Gamma^P}$ .

\setcounter{equation}{0}
\section{Results}

In  this section we present  closed  expressions for  the NCR obtained
from the one-loop proper PT vertex $\widehat{\Gamma}^{\mu}_{\gamma \nu
\bar{\nu}}$ given in Eq.(\ref{nuevo}), as well as the slope at $q^2=0$
of the   effective  self-energy $\widehat{\Pi}_{\mu\nu}^{\gamma {\chic
Z}}(q)$ defined  in  Eq.(\ref{hatPi}).  The  latter  quantity has been
traditionally considered as part of the NCR ;  however recently it has
been realized
\cite{PRW} that
it can instead be interpreted as part of
the effective mixing angle $sin^2_{\chic W, eff}(q^2)$.
This quantity, together with the effective charge
originating from the $ZZ$ propagator, contains the
universal (process-independent) corrections of every
neutral current process. 
At the theoretical level the separation appears to be
rather appealing, and the experimental disentanglement of 
these two contributions will be a matter of further study. 

We begin with the  transverse
PT self-energy $\widehat{\Pi}_{\mu\nu}^{\gamma {\chic Z}}(q)$;
it may be written in the form
\be
\widehat{\Pi}_{\mu\nu}^{\gamma {\chic Z}}(q) = \Bigg( g_{\mu \nu}
- \frac{q_{\mu} q_{\nu}}{q^2} \Bigg) 
{\widehat{\Sigma}}^{\gamma {\chic Z}} (q)\, .
\label{d4}
\ee
The scalar self-energy
$\widehat{\Sigma}^{\gamma {\chic Z}}(q)$ consists  
of a fermionic and a bosonic part,
i.e.
\be
{\widehat{\Sigma}}^{\gamma {\chic Z}} (q) = 
{\widehat{\Sigma}}^{\gamma {\chic Z}}_f (q) + 
{\widehat{\Sigma}}^{\gamma {\chic Z}}_b (q) \, ,
\label{d5}
\ee
\noindent
given by
\be
\displaystyle
{\widehat{\Sigma}}^{\gamma {\chic Z}}_f (q) = -\, 
\frac{\alpha}{6 \pi} \sum_f N^c_f (-Q_f) [C^{\chic R}_f + C_f^{\chic L}] 
\displaystyle
\Bigg[ \frac{1}{3} q^2 + 2 m_f^2 B_0 (0; m_f^2, m^2_f) -
(q^2 + 2m^2_f) B_0 (q^2; m_f^2, m_f^2) \Bigg] \, ,
\label{d6}
\ee
where
\be
\alpha = \frac{e^2}{4 \pi}\ \  ,\ \  
C^{\chic R}_f = - \frac{s_{\chic W}}{c_{\chic W}} Q_f\ \ ,\ \ 
C^{\chic L}_f = \frac{T^{\chic F}_{z} - s^2_{\chic W} Q_f}
{s_{\chic W} c_{\chic W}}\ \ ,\ \ N_f^c = 
\left\{ \begin{array}{l}
3 \; \hbox{quarks}  \\
1 \; \hbox{leptons} \end{array} \right. ,
\label{d7}
\ee
and 
\bea
{\widehat{\Sigma}}^{\gamma \chic{Z}}_b (q) &=& 
\frac{\alpha}{12 s_{\chic W} c_{\chic W} \pi} \Bigg\{ 
\Bigg[ \Bigg( 21 c^2_{\chic W} +  \frac{1}{2} \Bigg) q^2 + 
(12 c^2_{\chic W} - 2) M^2_{\chic W} \Bigg] 
B_0 (q^2; M^2_{\chic W}, M^2_{\chic W})
\nonumber\\
&&- (12 c^2_{\chic W} - 2) M^2_{\chic W}\, 
B_0(0; M^2_{\chic W}, M^2_{\chic W}) 
+ \frac{1}{3} q^2 \Bigg\} \, .
\label{d8}
\eea
$B_0$ is the standard Passarino-Veltman two-point function.
The counterterm corresponding to 
${\widehat{\Sigma}}^{\gamma {\chic Z}} (q)$ is
given by
\bea
\displaystyle
\frac{1}{2}\, \delta Z_{\gamma {\chic Z}} &=& - \Re e 
\frac{{\widehat{\Sigma}}^{\gamma {\chic Z}} (M_{\chic Z}^2)}{M_{\chic Z}^2}
\nonumber\\
&=& \frac{\alpha}{4 \pi}
\Bigg\{ \frac{2}{3} \sum_f N^c_f (-Q_f) [C^{\chic R}_f + C_f^{\chic L}]
\Bigg[\frac{1}{3} - (1 + 2 c_{\chic W}^2 x_f) \beta_f -
B_0 (0; m_f^2, m_f^2,) \Bigg] 
\nonumber\\
\displaystyle
&& \frac{- 1}{3 s_{\chic W} c_{\chic W}} \Bigg[ \frac{1}{3}
+ \Bigg( 21 c^2_{\chic W} +  \frac{1}{2} \Bigg)
B_0 (0; M^2_{\chic W}, M^2_{\chic W}) + \Big( 12 c_{\chic W}^4 
+ 19 c_{\chic W}^2 + \frac{1}{2} \Big) \beta_b \Bigg] \Bigg\}\, ,
\label{d9}
\eea
where 
\bea
x_f &=& \frac{m_f^2}{M_{\chic W}^2}\nonumber\\
\beta_f &=& 2\, \Bigg\{ 1 - \sqrt{4 c_{\chic W}^2 x_f - 1}\, 
\tan^{-1} \Bigg( \frac{1}{\sqrt{4 c_{\chic W}^2 x_f - 1}} \Bigg) \Bigg\}
\nonumber\\
\beta_b &=& 2\, \Bigg\{ 1 - \sqrt{4 c_{\chic W}^2 - 1}\, 
\tan^{-1} \Bigg( \frac{1}{\sqrt{4 c_{\chic W}^2 - 1}} \Bigg) \Bigg\}
\label{d10}
\eea
Thus,
the renormalized derivative at the origin is given by 
\be
\begin{array}{c}
\displaystyle
- \frac{{{\widehat{\Sigma}}^{\gamma {\chic Z}'}_{\chic R}(0)}}{M_{\chic Z}^2}
= \frac{3 G_{\chic F}}{2 \pi^2 \sqrt{2}}
\Bigg\{ \frac{2}{3} \sum_f N^c_f (-Q_f) [C^{\chic R}_f + C_f^{\chic L}]
\Bigg[s_{\chic W} \Bigg(- \frac{1}{3} 
+ [1 + 2 c_{\chic W}^2 x_f] \beta_f \Bigg) \Bigg] 
\\[0.5cm]
\displaystyle
+ \frac{1}{3 c_{\chic W}} \Bigg[- \Big( 2c_{\chic W}^2 - \frac{1}{3}\Big) 
+ \Big( c_{\chic W}^2 (12 c_{\chic W}^2 + 19) + \frac{1}{2} \Big) \beta_b 
\Bigg] \Bigg\}\, .
\end{array}
\label{d11}
\ee
where $G_{\chic F}$ is the Fermi constant.
Evaluating this last expression, summing over  
all quarks and leptons, we obtain
\be
- \frac{{{\widehat{\Sigma}}^{\gamma {\chic Z}'}_{\chic R}(0)}}{M_{\chic Z}^2}
= 1.6 \times 10^{-32} \ cm^2.
\label{d12}
\ee

Next we turn to the proper vertex.
The Dirac form-factor $F_1(q^2)$ 
is usually defined as the cofactor of the
$\gamma^{\mu}$ 
in the Lorentz decomposition of
$\widehat{\Gamma}^{\mu}_{\gamma \nu \bar{\nu}}$.
An elementary calculation of the two diagrams contributing to 
$\widehat{\Gamma}^{\mu}_{\gamma \nu \bar{\nu}}$ yields 
\bea
\displaystyle
F_1 (q^2) &=& -\, \frac{\alpha e}{8 \pi s_{\chic W}^2} 
\Bigg\{1 + \Bigg( \frac{1}{2} + \frac{M^2_{\chic W}}{q^2} \Bigg)
\Big[ B_0 (q^2; m_l^2, m_l^2) 
- B_0 (q^2; M^2_{\chic W}, M_{\chic W}^2) \Big]
\nonumber\\[0.5cm]
\displaystyle
&& +\, M_{\chic W}^2 \Bigg( 2 + \frac{M^2_{\chic W}}{q^2} \Bigg)
C_0 (0, q^2, 0; m_l^2, M_{\chic W}^2, M^2_{\chic W})
+\, \frac{(q^2 + M_{\chic W}^2)^2}{q^2}
C_0 (0, q^2, 0; M^2_{\chic W}, m_l^2, m_l^2)
\Bigg\}\, ,\nonumber\\
&& {}
\label{d1}
\eea
where $C_0$ is the Passarino-Veltman
three-point function,
and $m_{\chic l}$ is the
mass of the charged isodoublet
partner of the neutrino under consideration.

In the limit $q^2 \rightarrow 0$, 
$F_1 = - \frac{q^2}{6} <r^2_\nu>_1$
with
\be
\big <r^2_\nu\,  \big>_1 =\, 
\frac{G_{\chic F}}{4\, {\sqrt 2 }\, \pi^2} \Bigg[3 - 2 \log 
\Bigg(\frac{m_{\chic l}^2}{M_{\chic W}^2} \Bigg) \Bigg]
\label{ncr}
\ee
The numerical evaluation of the above expression for
the three different neutrino species yields:
\be
\big< r^2_\nu\, \big>_1 = 
\left\{ 
\begin{array}{l}
4.1 \times 10^{- 33}\ cm^2 \ \ \  \hbox{for $\nu_e$}  \\
2.4 \times 10^{- 33}\ cm^2 \ \ \ \hbox{for $\nu_\mu$} \\
1.5 \times 10^{- 33}\ cm^2 \ \ \ \hbox{for $\nu_\tau$} 
\end{array} \right. .\\
\label{d3}
\ee

For a spin $\frac{1}{2}$ particle, like the neutrino, the electric 
Sachs form-factor $G_{\chic E} (q^2)$ is given in terms of the Dirac and
Pauli form-factors as 
\be
G_{\chic E} (q^2) = F_1 (q^2) + \frac{q^2}{4m^2_{\nu}} F_2 (q^2)
\ee
with $F_1 (0)=0$ and $F_2 (0) = \mu_{\nu}$, where $\mu_{\nu}$
is the anomalous magnetic moment of the neutrino. To leading order
in the mass of the Dirac neutrino, the value of 
$\mu_{\nu}$ is given by \cite{N5}
\be
\mu_{\nu} = \frac{3 m^2_{\nu} G_{\chic F}}{4\, {\sqrt 2 }\, \pi^2} 
\, .
\ee
Its contribution (the so-called Foldy term \cite{Foldy}) to the
charged radius is thus
\bea
\big< r^2_\nu\, \big>_2 &=& - 
\frac{9 G_{\chic F}}{8\, {\sqrt 2 }\, \pi^2} \nonumber\\ 
&=& - 3.7 \times 10^{- 34}\ cm^2
\eea

\setcounter{equation}{0}
\section{Conclusions and Outlook}

\noindent

This   work has addressed  the  problem of correctly   defining in the
context  of  the Standard  Model  the one-loop neutrino charge radius.
The discussion of  this topic has a long  history and it  was attacked
along its path with  the philosophy of  making ``proposals''.  In this
work we   have   presented a definite   solution  to  this problem  by
resorting  to  the well-defined electroweak gauge-invariant separation
of a physical amplitude into ``{\it effective}'' (as opposed to ``{\it
diagrammatic}'')   self-energy,  vertex    and box     descompositions,
implemented within  the  pinch technique formalism.   These  effective
Green's  function are   completely independent   of the   gauge-fixing
parameter regardless of the  gauge-fixing scheme chosen.  This latter
property  has  been demonstrated  explicitly by  considering  two very
different   gauge-fixing  schemes,   i.e.    the renormalizable  gauge
$R_\xi$-scheme  and the  background  field method.   In  addition, the
pinch technique ``{\it effective}''  Green's functions satisfy simple,
QED-like  Ward identities  which  replace the involved  Slavnov-Taylor
identities.  In  the course  of the derivation,   we have explained in
detail  that  the  solution  of  the  problem   relies on  the  proper
identification of the    longitudinal degrees   of freedom, and     in
particular those  associated  with the  non-abelian  character of  the
theory (tri-linear vertices).  In addition, we  have elaborated on the
tree-level  origin of the  one-loop cancellation mechanism, especially
its realization in the absence of  $t$-channel tree-level graphs, i.e.
in the case of $W$-pair production by right-handed target fermions.

The  conceptual requirement that  the effective electromagnetic vertex
of a particle has to be process-independent, i. e., independent of the
target used to probe the properties of  the particle, is automatically
implemented in our  solution.   In fact, the  right-handedly polarized
electrons provide the   dynamical implementation of  the earlier claim
\cite{JPPT} that the effective  
boxes should  not   enter in  the  definition of  the neutrino  charge
radius.   When  studying  the neutrino   charge form-factor, we  have
compared   the  effective neutrino  vertex obtained   as  seen by both
unpolarized  electrons and right-handedly  polarized  electrons. As it
has to be, the  effective vertex is  the  same.  However, in terms  of
diagrams  this  common  result  is   obtained  from vastly   different
contributions in both cases: in  particular, for unpolarized electrons
there is a $WW$ box contribution to the $S$-matrix, which is absent in
the  latter case.  We  have  thus a neutrino  charge  radius which  is
independent of the gauge  fixing parameter or the gauge-fixing scheme,
couples  electromagnetically to the  external probe  and is completely
independent of it.   The  resulting one-loop  expression for the   NCR
given in  Eq.(\ref{ncr}) depends only   on the mass of  the associated
charged lepton.

Equipped  with  this theoretical identification,   one can address the
question of  the  observable character of  each  one of the  separated
effective contributions to a given process.  A positive answer to this
question would imply that the sub-amplitudes thus constructed not only
are endowed with nice theoretical properties but are in fact physical.
For  example, it  is  known that the electroweak effective
charge can be  reconstructed from specific, separable contributions of
the physical differential cross-section describing $W$-pair production
\cite{PRW}.
A  similar study is  now in  progress for the  neutrino charge radius,
making  use of the opportunities offered  by neutrino and antineutrino
cross-sections   and by  the   different   neutrino flavours   to take
appropriate ratios.

It goes without saying that the study presented here for the neutrino
charge radius can be extended to all $q^2$-values (either space-like or
time-like) of the charge form-factor, and to all the other form-factors 
of the electromagnetic effective vertex.

\section*{Acknowledgments}

\noindent

L.G.C.R. is indebted to the Universidad de Valencia, for his fellowhship.
This research was supported by CICYT, Spain, under Grant AEN-99/0692.

\end{document}